\documentstyle[psfig,times]{jfm}

\begin{document}
\title
[An Improved Exact Riemann Solver for Multidimensional
Relativistic Flows]
{An Improved Exact Riemann Solver for Multidimensional
Relativistic Flows}

\author
[L. Rezzolla, O. Zanotti, \& J.A. Pons]
{LUCIANO REZZOLLA$^{1,2}$,  OLINDO ZANOTTI$^1$
\and JOSE A. PONS$^3$}

\affiliation{$^1$~SISSA, International School for Advanced Studies,
	Trieste, Via Beirut 2--4, 34014 Trieste, Italy\\[\affilskip]
	     $^2$~INFN, Physics Department, University
	of Trieste, Via Valerio 2, 34127 Trieste, Italy\\[\affilskip]
	     $^3$~Physics Department, University
	of Rome ``La Sapienza'', Piazzale Aldo Moro 2, I-00185 Rome, Italy} 

\date{\today}

\maketitle

\begin{abstract}
We extend our approach for the exact solution of the
Riemann problem in relativistic hydrodynamics to the case
in which the fluid velocity has components tangential to
the initial discontinuity. As in one-dimensional flows,
we here show that the wave-pattern produced in a
multidimensional relativistic Riemann problem can be
predicted entirely by examining the initial
conditions. Our method is logically very simple and
allows for a numerical implementation of an exact Riemann
solver which is both straightforward and computationally
efficient. The simplicity of the approach is also
important for revealing special relativistic effects
responsible for a smooth transition from one wave-pattern
to another when the tangential velocities in the initial
states are suitably varied. While the content of this
paper is focussed on a flat spacetime, the local Lorentz
invariance allows its use also in fully general
relativistic calculations.
\end{abstract}

\section{Introduction}

	Since the solution of the Riemann problem has
been introduced in numerical hydrodynamics (Godunov
1959), high resolution shock capturing methods have
become a common tool for handling nonlinear
hydrodynamical waves in a variety of physical situations
and the subject of detailed mathematical analyses both in
Newtonian (Leveque 1992, Toro 1997) and relativistic
regimes (Smoller \& Temple 1993). Nonlinear waves of this
type are very common in astrophysical scenarios (such as
gamma-ray bursts, accretion onto compact objects,
relativistic jets, supernova explosions) in which the
motion of the fluid is characterised by relativistic
speeds and by the appearance of strong discontinuities
(see Iba\~nez \& Mart\'{\i}, 1999 and Mart\'{\i} \&
M\"uller, 1999 for a review). In a recent paper, Rezzolla
\& Zanotti (2001, hereafter paper I) have proposed a new
procedure for the exact solution of the Riemann problem
which uses the relativistically invariant relative
velocity between the unperturbed ``left'' and ``right''
states of the fluid in order to extract important
information contained in the initial data and that more
traditional approaches were not able to put into
evidence. Most notably, it was shown that, given a
Riemann problem with assigned initial conditions, it is
possible to determine in advance both the wave-pattern
that will be produced after the removal of the initial
planar discontinuity and the bracketing interval of the
unknown pressure in the region that forms behind the wave
fronts. Besides clarifying some aspects of the Riemann
problem, the approach proposed by Rezzolla \& Zanotti,
and that was foreseen by Landau \& Lifshitz (1959) in
Newtonian hydrodynamics, was shown to be computationally
more efficient as compared to the more traditional ones.

	In a recent work, Pons {\it et al} (2000) have
extended the analytic solution of the one-dimensional
Riemann problem (Mart\'{\i} \& M\"uller 1994) to the case
in which nonzero velocities tangential to the initial
planar discontinuity are present. An important result
obtained by Pons {\it et al} (2000) was to show that the
introduction of tangential velocities and the appearance
of global Lorentz factors linking quantities on either
side of the discontinuity can affect the solution of the
Riemann problem considerably. In the present paper we
show that the approach by Rezzolla \& Zanotti can be
successfully extended to this more general case and with
the same advantages that were found in the case of zero
tangential velocities. Within this new procedure, the
introduction of tangential velocities is imprinted in the
expression for the jump of the velocity normal to the
discontinuity surface, and this has allowed us to reveal
interesting special relativistic effects. In relativistic
hydrodynamics, in fact, the wave-pattern produced by the
decay of the initial discontinuity can be changed by
simply varying the tangential velocities in the initial
states, while keeping the rest of the thermodynamic
quantities of the Riemann problem unmodified.  This
effect has no analogue in Newtonian hydrodynamics
(Rezzolla \& Zanotti 2002).

	The plan of the paper is as follows: after a
review of the method in Section 2, we report in Section 3
the hydrodynamical equations relevant for the present
discussion of nonzero tangential velocities. In Section 4
we show how to use the invariant relative velocity to
extract from the initial data the information on the
wave-pattern produced, while Section 5 is devoted to the
presentation of the special relativistic effects. The
conclusions of the paper are in Section 6, and two
Appendices complete the discussion providing the
mathematical details of the results obtained in the main
text.

	We use in this paper a system of units in which
$c = 1$, a space-like signature $(-,+,+,+)$ and a
Cartesian coordinate system $(t,x,y,z)$. Greek indices
are taken to run from 0 to 3, Latin indices from 1 to 3
and we adopt the standard convention for the summation
over repeated indices.

\section{A brief review of the method}
\label{s:eqs}

	In a flat spacetime consider a perfect fluid
described by the stress-energy tensor:
\begin{equation}
T^{\mu\nu}\equiv (e+p)u^\mu u^\nu+p \eta^{\mu\nu}
	= \rho h u^\mu u^\nu+p \eta^{\mu\nu} \ ,
\end{equation}
where $\eta^{\mu\nu}={\rm diag}(-1,1,1,1)$ and $e$, $p$,
$\rho$, and $h$ are the proper energy density, the
isotropic pressure, the proper rest mass density, and the
specific enthalpy, respectively.

	Further assume the fluid to consist of an initial
``left'' state (indicated with an index $1$) and an
initial ``right'' state (indicated with an index $2$),
each having prescribed and different values of uniform
pressure, rest-mass density and velocity. The two
discontinuous states are initially separated by a planar
surface $\Sigma_0$ placed at a constant value of the $x$
coordinate so that the unit space-like 4-vector
$n^{\mu}_0$ normal to this surface at $t = 0$ has
components $n^{\mu}_0\equiv(0,1,0,0)$\footnote{More
precisely, the unit normal to the hypersurface $\Sigma_0$
is defined as the one-form mapping each vector tangent to
the surface into zero.}. Because we are considering a
multidimensional flow, the fluid 4-velocity on either
side of the initial discontinuity is allowed to have
components in spatial directions orthogonal to
$n^{\mu}_0$, i.e.
\begin{equation}
u^{\mu} \equiv W(1,v^x,v^y,v^z) \ ,
\end{equation}
where $W^2 = (1-v^2)^{-1}$ is the square of the Lorentz
factor and $v^2 \equiv v^i v_i= (v^x)^2 + (v^y)^2 +
(v^z)^2$ is the norm of the 3-velocity. The
hydrodynamical properties of the initial left and right
states are described by the ``state-vectors''
\begin{eqnarray}
{\mathbf U}_{1,2}& =& \left(\begin{array}{c}
p\\
\rho\\
v^x\\
v^t
\end{array}\right)_{1,2}
\nonumber \ ,
\end{eqnarray}
where we have indicated with $v^t\equiv[(v^y)^2 +
(v^z)^2]^{1/2}$ the {\em tangential} component of the
three velocity, satisfying the obvious relativistic
constraint that $(v^t)^2+(v^x)^2 \le 1$ (Note that in
paper I the 4-velocity vector was chosen so that
$v^y=v^z=0=v^t$).  Hereafter, we will refer to $v^x$ as
the {\em normal} velocity.

	The fluid states ${\mathbf U}_1$ and ${\mathbf
U}_2$ represent the initial conditions of a
multidimensional ```Riemann problem'' whose solution
consists of determining the flow that develops when the
system is allowed to relax. In general, the temporal
evolution can be indicated as (Mart\'{\i} \& M\"uller
1994)
\begin{equation}
\label{schema}
L{\cal W}_\leftarrow L_*{\cal C}R_*{\cal W}_\rightarrow R
	\ ,
\end{equation}
where ${\cal W}$ denotes a nonlinear wave (either a
shock, ${\cal S}$, or a rarefaction wave, ${\cal R}$),
propagating towards the left $(\leftarrow)$ or the right
$(\rightarrow)$ with respect to the initial
discontinuity. Moreover, $L_*$ and $R_*$ are the new
hydrodynamic states that form behind the two nonlinear
waves propagating in opposite directions. A contact
discontinuity, ${\cal C}$, separates the region $L_*$ and
$R_*$, and it is characterised by the fact that both the
pressure and the normal velocity are continuous across
it, while both the rest-mass density and the tangential
velocities can be discontinuous.

	The approach introduced in paper I focuses on
$(v^x_{12})_0$, the relativistic invariant expression for
the relative velocity between the two unperturbed initial
states.  By construction, this quantity measures the
relativistic jump of the fluid velocity normal to the
discontinuity surface.  The solution of the relativistic
Riemann problem is then found after the pressure in the
region between the two nonlinear waves, $p_*$, is
calculated as the root of the nonlinear equation
\begin{equation}
\label{our}
v^x_{12}(p_*) - (v^x_{12})_0 = 0 \ ,
\end{equation}
where $v^x_{12}(p_*)$ has a functional form that is {\it
different} for each of the {\it three} possible
wave-patterns that might result from the decay of the
initial discontinuity. The key aspect of the new approach
is that the wave-pattern produced by the decay of the
initial discontinuity can be entirely predicted in terms
of the initial data ${\bf U}_{1,2}$. This represents an
important advantage since it allows to deduce in advance
which set of equations to use for the solution of the
exact Riemann problem and the interval bracketing the
root of (\ref{our}).

	The validity of the approach discussed in paper I
is in the mathematical proof that the function
$v^x_{12}=v^x_{12}(p_*)$ is monotonically increasing with
$p_*$ and it is composed of three branches corresponding
to the three possible wave-patterns. Furthermore, the
three different branches always join smoothly through
specific values of $v^x_{12}(p_*)$, denoted respectively
as $({\tilde v^x}_{12})_{_{2S}}$, $({\tilde
v^x}_{12})_{_{SR}}$, $({\tilde
v^x}_{12})_{_{2R}}$\footnote{Here $2S$, $SR$ and $2R$
indicate a two shock, a shock-rarefaction and a two
rarefaction wave-pattern, respectively.}  (cf. Fig. 1 of
paper I). In view of these properties, it is possible to
compare $(v^x_{12})_0$ with the relevant limiting values
$({\tilde v^x}_{12})_{_{2S}}$, $({\tilde
v^x}_{12})_{_{SR}}$, $({\tilde v^x}_{12})_{_{2R}}$
constructed from the initial conditions and determine,
prior to the solution of equation (\ref{our}), both the
wave-pattern that will be produced and the functional
form of $v^x_{12}(p_*)$ to be used. The logical scheme to
be followed in this procedure is synthesised as follows
\begin{equation}
\begin{array}{rlc}
\label{logic}
\\ (v^x_{12})_0 > ({\tilde v^x}_{12})_{_{2S}}: & \ \
L{\cal S}_\leftarrow L_*{\cal C}R_*{\cal S}_\rightarrow R
& v^x_{12}(p_*)=(v^x_{12})_{_{2S}}\\
({\tilde v^x}_{12})_{_{SR}}< (v^x_{12})_0  \leq  ({\tilde
	v}_{12})_{_{2S}}: &  
	\ \ L{\cal R}_\leftarrow L_*{\cal
	C}R_*{\cal S}_\rightarrow R, \ \ & 
	v^x_{12}(p_*)=(v^x_{12})_{_{SR}}\\ 
({\tilde v^x}_{12})_{_{2R}}< (v^x_{12})_0\leq ({\tilde
	v}_{12})_{_{SR}}: &  
	\ \ L{\cal R}_\leftarrow L_*{\cal C}R_*
	{\cal R}_\rightarrow R,\ \ &  
	v^x_{12}(p_*)=(v^x_{12})_{_{2R}}\\ 
(v^x_{12})_0 \leq  ({\tilde v^x}_{12})_{_{2R}}: &  
	\ \ L{\cal R}_\leftarrow L_*{\cal C}R_*
	{\cal R}_\rightarrow R~~{\rm with\ vacuum},\ \ &
	\ - \\ \\
\end{array}
\end{equation}
After the initial relative normal-velocity $(v^x_{12})_0$
has been compared with the limiting values for the
different branches and the correct wave-pattern [i.e. the
functional form $v^x_{12}(p_*)$] has been found, the
numerical solution of the exact Riemann problem can be
performed. Furthermore, the knowledge of the wave-pattern
relevant for the solution of the problem provides the
values $p_{\rm max}$ and $p_{\rm min}$ bracketing the
unknown pressure $p_*$. In the case of a one-dimensional
flow, the bracketing interval is sinthesized in the
scheme below

\begin{center}
\vskip 0.5truecm
\begin{tabular}{|l|c|c|c|}
\cline{1-4} & & &\\ 
&$L{\cal S}_\leftarrow L_*{\cal C}R_*{\cal S}_\rightarrow R$&
 $\ \ L{\cal R}_\leftarrow L_*{\cal C}R_*{\cal S}_\rightarrow R \ \ $ 
&$\ \ L{\cal R}_\leftarrow L_*{\cal C}R_*{\cal R}_\rightarrow R\ \ $
\\ 
\cline{1-4}  & & &\\ 
 $p_{\rm min}$		& max$(p_1,p_2)$	&min$(p_1,p_2)$
			& 0
\\
\cline{1-4}  & & &\\ 
$p_{\rm max}$		& $\infty$		&max$(p_1,p_2)$
			&min$(p_1,p_2)$
\\ \cline{1-4} 
\end{tabular}
\vskip 0.8truecm
\end{center}
	In practice, this novel approach has two major
advantages. Firstly, it results into a straightforward
numerical implementation and in a more efficient
numerical algorithm with a reduction of the computational
costs. Secondly, it simplifies considerably the logic
formulation of the relativistic Riemann problem. As we
shall discuss in Section \ref{effect}, this feature has
been essential in revealing new special relativistic
effects.

	The extension of the approach presented in paper
I to the case when tangential velocities are present is
straightforward and equation (\ref{our}) as well as the
logical scheme (\ref{logic}) apply unmodified\footnote{
Note that in the case of two shock waves propagating in
opposite directions the maximum value for the pressure
needs not be infinite but can be determined from the
solution of the equation $v^x_{12}(p_*) - (v^x_{12})_{\rm
max} = 0$, where $(v^x_{12})_{\rm max}$ is the maximum
allowed value of the normal relative velocity that is
compatible with the assigned initial tangential
velocities.}.  
The only changes introduced by the
presence of tangential velocities are restricted to the
expressions for the limiting values of the relative
velocity $({\tilde v^x}_{12})_{_{2S}}$, $({\tilde
v^x}_{12})_{_{SR}}$, $({\tilde v^x}_{12})_{_{2R}}$. The
details of these changes will be presented in the
following Sections which have been written for a generic
equation of state (EOS) and use an ideal fluid EOS as a
test case.

\section{Hydrodynamical Relations across the waves}
\label{intro}
	
	As discussed in the previous Section, the
expression for the relative normal-velocity between the
two initial states of the Riemann problem represents the
building block in our approach and, to simplify our
notation, hereafter we will refer to the different flow
regions using the following mapping
\begin{equation}
\label{mapping}
L{\cal W}_\leftarrow L_*{\cal C}R_*{\cal W}_\rightarrow R
\hskip 1.0 truecm \Longleftrightarrow \hskip 1.0 truecm 
1\; {\cal W}_\leftarrow  \; 3\; {\cal C}\;
3'\;{\cal W}_\rightarrow \; 2
\ .
\end{equation}

	While the values of $v^x_{12}$ are relativistic
invariants, there exists a reference frame which is
better suited to evaluate this quantity. In the reference
frame of the contact discontinuity, in fact, the normal
velocities behind the nonlinear waves are, by definition,
zero (i.e. $v^x_{3,{\cal C}} = 0 = v^x_{3',{\cal C}}$)
and the relative velocities across the nonlinear waves
measured in this reference frame will be
\begin{eqnarray}
(v^x_{13})_{,\cal C}
	&\equiv&
	\frac{v^x_{1,{\cal C}}  - v^x_{3,{\cal C}}}
	{1-  (v^x_{1,{\cal C}})  (v^x_{3,{\cal C}})}
	= v^x_{1,{\cal C}} \ , \\
(v^x_{23'})_{,\cal C}
	&\equiv&
	\frac{v^x_{2,{\cal C}} - v^x_{3',{\cal C}}}
	{1 - (v^x_{2,{\cal C}}) (v^x_{3',{\cal C}})}
	=v^x_{2,{\cal C}}  \ .
\end{eqnarray}
Because of their invariance, the normal velocity jumps
across the nonlinear waves measured in the Eulerian frame
can be expressed as
\begin{eqnarray}
\label{1c}
(v^x_{13}) &=&  
	\frac{v^x_{1}  - v^x_{3}}
	{1-  v^x_{1}  v^x_{3}} = (v^x_{13})_{,\cal C} = 
	v^x_{1,{\cal C}} \ , \\
\label{2c}
(v^x_{23'}) &=&  
\frac{v^x_{2}  - v^x_{3'}}
	{1-  v^x_{2}  v^x_{3'}}
	= (v^x_{23'})_{,\cal C}
	= v^x_{2,{\cal C}}  \ .
\end{eqnarray}
As a result, the relative normal-velocity between the two
initial states can be written as
\begin{equation}
\label{g_vrel}
v^x_{12} = (v^x_{12})_{,{\cal C}} = 
	\frac{v^x_{1,{\cal C}} - v^x_{2,{\cal C}}}
	{1-  (v^x_{1,{\cal C}}) (v^x_{2,{\cal C}})} \ .
\end{equation}

In what follows we will briefly discuss how to calculate
the normal-velocity jump across a shock wave and a
rarefaction wave, respectively. The expressions derived
in this way will then be used to calculate $v^x_{1,{\cal
C}}$ and $v^x_{2,{\cal C}}$ necessary to build
$v^x_{12}=v^x_{12}(p_*)$ [cf. eq. (\ref{g_vrel})].

\subsection{Jumps Across a Shock Wave}

	Following Pons {\it et al} (2000), we use the
Rankine-Hugoniot conditions in the fixed Eulerian
reference frame to express the normal velocity of the
fluid on the back of the shock front in terms of the
pressure\footnote{Note that when tangential velocities
are present, calculating jump conditions in the rest
frame of the shock front is not particularly
advantageous. In this case, in fact, the velocity jump
across the shock cannot be expressed as an algebraic
relation among the thermodynamical quantities across the
shock [cf. eq. (3.1) of Rezzolla \& Zanotti
2001]. Rather, the ratio of the velocities ahead and
behind the shock front need to be found as root of a
nonlinear equation. Analytic solutions to this equation
can be found only in the {\it weak-shock} limit or for an
ultrarelativistic equation of state (see Koenigl 1980 for
a discussion).}
\begin{equation}
\label{v_b_s}
v_b^x = \frac{h_a W_a v_a^x + W_s(p_b - p_a)/J}{h_a W_a +
	(p_b - p_a)[W_s v_a^x/J + 1/(\rho_a W_a)]}
	\ ,
\end{equation}
where $J$ is mass flux and the subscripts $b$ and $a$
denote a quantity evaluated on the back or ahead of the
wave front, respectively. In equation (\ref{v_b_s}), $W_s
\equiv (1- V^2_s)^{-1/2}$ is the Lorentz factor of the
shock velocity $V_s$, with the latter being
\begin{equation}
\displaystyle{
V_s^{\pm} = \frac{ \rho_a^2 W_a^2 v^x_a  \pm   
	|J| \sqrt{J^2 + \rho_a^2 W_a^2 [1 -(v_a^x)^2]}}
            { \rho_a^2 W_a^2 + J^2 } }  \ ,
\label{velshock}
\end{equation}
and with the $\pm$ signs referring to a shock wave
propagating to the right or to the left, respectively.

	Besides giving the jump in the normal velocity
across a shock wave, expression (\ref{v_b_s}) states that
the two regions of the flow across the shock wave are
effectively coupled through a Lorentz factor which, we
recall, is built also in terms of the tangential
velocities. This is a purely relativistic feature and an
important difference from Newtonian hydrodynamics, in
which the solution of the Riemann problem does not depend
on the tangential component of the flow. Many of the
consequences introduced by this coupling will be further
discussed in Section~\ref{effect} but one of them can
already be appreciated when looking at the jumps across
the shock wave of the tangential velocities that can be
written as
\begin{equation}
\label{rh_tan}
[[ h W v^y]] = 0 = [[ h W v^z]] \ , 
\end{equation}
and where we have adopted the standard notation in which
$[[F]] \equiv F_a - F_b$ (Anile 1989). Conditions
(\ref{rh_tan}) basically state that the ratio $v^y/v^z$
remains unchanged through rarefaction waves and therefore
that the tangential velocity 3-vector does not rotate but
can change its norm. This property, which applies also
across rarefaction waves, represents a major difference
from the behaviour of the tangential velocities across
Newtonian nonlinear waves. In this latter case, in fact,
the tangential velocity vector does not change its norm,
nor rotates: $[[v^y]]= 0 =[[v^z]]$.

	The square of the mass flux across the wave can
be expressed as
\begin{equation}
\label{flux1}
J^2 = -\frac{[[p]]}{[[h/\rho]]} \ ,
\end{equation}
where the ratio $h/\rho$ in the shocked region can be
calculated through the Taub adiabat (Taub, 1948)
\begin{equation}
\label{Taub1}
[[h^2]] =
\left(\frac{h_a}{\rho_a}+\frac{h_b}{\rho_b}\right)[[p]] \ .
\end{equation}

	In a general case, the mass flux can be obtained
as a function of just one thermodynamical variable
($p_*$) after using the EOS and the physical ($h \ge 1$)
solution of the nonlinear equation (\ref{Taub1}). In the
case of an ideal fluid EOS,
\begin{equation}
\label{pol_eos}
p = (\gamma-1)\rho\epsilon = k(s)\rho^{\gamma} ,
\end{equation}
where $\gamma$ is the adiabatic index, and $k(s)$ is the
polytropic constant (dependent only on the specific
entropy $s$), this can be done explicitly because
(\ref{flux1}) and (\ref{Taub1}) take respectively the
simple form
\begin{equation}
\label{flux2}
J^2=-\frac{\gamma}{\gamma-1}\frac{[[p]]}{[[h(h-1)/p]]} \ , 
\end{equation}
and
\begin{equation}
\label{Taub2}
\left[1+\frac{(\gamma-1)(p_a-p_b)}{\gamma p_b}\right]h_b^2
	-\frac{(\gamma-1)(p_a-p_b)}{\gamma p_b}h_b 
	+\frac{h_a (p_a-p_b)}{\rho_a}-h_a^2 = 0 \ .
\end{equation}

\subsection{Jumps Across a Rarefaction Wave}

	When considering a rarefaction wave it is
convenient to introduce the self-similar variable\break
\hbox{$\xi \equiv x/t$} in terms of which similarity
solutions to the hydrodynamical equations can be
found. An explicit expression for $\xi$ can be obtained
by requiring that non-trivial similarity solutions for
the rarefaction wave exist. This then yields (see Pons
{\it et al} 2000 for details)
\begin{equation}
\label{xi6} 
\xi  =  \frac{v^x(1-c_s^2) \pm c_s
	\sqrt{(1-v^2)[1 - v^2 c_s^2 -
	(v^x)^2(1-c_s^2)]}}{1-v^2c_s^2} \ ,
\end{equation}
where here too the $\pm$ signs correspond to rarefaction
waves propagating to right or to the left, respectively.
In the case of a perfect fluid, the isentropic character
of the flow allows to calculate the velocity on the back
of the wave as a solution of an ordinary differential
equation
\begin{equation}
\label{v_b_r}
\frac{dv^x}{dp}= \frac{1}
       {\rho h W^2}\frac{(1 - \xi v^x)} {(\xi - v^x)} \ ,
\end{equation}

	In principle, to calculate the normal fluid
velocity at the tail of the rarefaction wave one should
solve the ordinary differential equation (\ref{v_b_r}),
which might be very expensive numerically. To overcome
this, it is convenient to make use of constraints such as
those in expressions (\ref{rh_tan}) (which remain valid
also across a rarefaction wave) and express equation
(\ref{v_b_r}) in a different way. Defining ${\cal A}
\equiv h_a W_a v^t_a = h_b W_b v^t_b$, the tangential
velocity along a rarefaction wave can be expressed as
\begin{equation}
\label{vt2}
(v^{t})^2 = {\cal A}^2 \left[ \frac{1-(v^{x})^2}{h^2 + 
	{\cal A}^2} \right]\ .
\end{equation}
This allows us to eliminate the dependence on $v^t$ from
equation (\ref{xi6}). From the definition of the Lorentz
factor and equation (\ref{vt2}) it is straightforward to
obtain
\begin{equation}
W^2 = \frac{h^2+{\cal A}^2}{h^2(1-(v^x)^2)}  \ ,
\end{equation}
and after some algebra one can arrive at
\begin{equation}
\label{o_m_xiv}
\frac{(1 - \xi v^x)} {(\xi - v^x) } = \pm 
	\frac{\sqrt{h^2 + {\cal A}^2(1-c_s^2)}}{h c_s}
\end{equation}
Using this results, equation (\ref{v_b_r}) can be written
as follows:
\begin{equation}
\label{raref2}
\frac{dv^x}{1-(v^{x})^2}= \pm \frac{\sqrt{h^2 + {\cal A}^2(1-c_s^2)}}
       {(h^2 + {\cal A}^2)} \frac{dp}{\rho~c_s} \ ,
\end{equation}
Note that in this way we have isolated the
thermodynamical quantities on the right hand side of
(\ref{raref2}) and the kinematical ones on the left hand
side, which can then be integrated analytically. For some
particular cases (for example when the sound speed is
constant), the right hand side too is integrable but for
a generic EOS a numerical integration is necessary. The
velocity at the tail of the rarefaction wave can then be
obtained directly as
\begin{equation}
v_b^x = \tanh{\cal B}~,
\label{newraref}
\end{equation}
where
\begin{equation}
{\cal B} \equiv \frac{1}{2}\log{ \left( \frac{1+v_a^x}{1-v_a^x} \right) }
	\pm \int_{p_a}^{p_*} \frac{\sqrt{h^2 + {\cal A}^2(1-c_s^2)}}
        {(h^2 + {\cal A}^2)} \frac{dp}{\rho~c_s} 
\label{newraref2}
\end{equation}
Here, $h=h(p,s)$, $\rho=\rho(p,s)$, and $c_s=c_s(p,s)$,
and the isentropic character of rarefaction waves allows
to fix $s=s_a$.  Despite its complicated look, the
integrand is a smooth, monotonic function of $p$, and a
Gaussian quadrature with only 10-20 points has proved to
be more accurate and efficient than a third order
Runge-Kutta integrator requiring hundreds of function
evaluations to solve (\ref{v_b_r}). Testing and details
on the numerical implementation of this procedure will be
reported in a separate work.

	A couple remarks should now be made about the
flow properties of a multidimensional and relativistic
Riemann problem. The first remark is about the flow
conditions at the contact discontinuity where, as
mentioned before, not only the rest-mass density but also
the tangential velocities can be discontinuous. It is
well known that under these conditions a Kelvin-Helmholtz
(or shear-layer) instability could develop
(Chandrasekhar, 1961), reach the nonlinear waves if these
move subsonically and thus destroy the solution of the
Riemann problem. While this could be a concern in
principle, it is never a problem in practice. The reason
for this is simple and is related to the fact that any
implementation of an exact Riemann solver in a numerical
code will provide the solution of the Riemann problem at
a new time-level which is ``Courant-Friedrichs-Lewy
limited'' from the previous one (Press {\it et al}
1986). Besides providing numerical stability, this
time-step limitation also guarantees that the solution at
the new time-level is everywhere causally connected with
the one at the previous one. As a result, even if a
Kelvin-Helmholtz instability would develop at the contact
discontinuity propagating outward in the flow at the
sound speed, this would not be able to influence the
dynamics of the nonlinear waves over the time of interest
in numerical calculations.

	The second remark is about the presence of
tangential velocities across a shock-wave which can be in
general discontinuous. While the stability of
relativistic planar shocks in one-dimensional flows has
been the subject of several investigations in the past
(Anile \& Russo 1987a, 1987b, Anile 1989), very little if
anything at all is known about the stability properties
of shock waves with discontinuous tangential
velocities. It is reasonable to expect that a nonzero
mass flux across these discontinuities would stabilise
them against the Kelvin-Helmholtz instability, but
further work needs to be done in this direction before a
conclusion can be drawn.

\section{Limiting relative velocities}
\label{urca}

	As mentioned in the previous Sections, the basic
operation in our approach consists of calculating the
relative normal-velocity across the two initial states
and comparing it with the limiting relative velocities
for each of the three possible wave-patterns. In
practice, this amounts to calculating equation
(\ref{g_vrel}) making use of expressions (\ref{1c}) and
(\ref{2c}). In the following we will briefly discuss the
guidelines for the evaluation of the limiting relative
velocities. In doing so we will adopt the convention of
paper I and assume that $p_1>p_2$, with the $x-$axis
normal to the discontinuity surface being positively
oriented from $1$ to $2$.

\subsection{$1\; {\cal S}_\leftarrow\; 3\; 
	{\cal C}\; 3'\; {\cal S}_\rightarrow\; 2$: 
	Two Shock Waves}
\label{case_i}

	We first consider a wave-pattern in which two
shock waves propagate in opposite directions. In this
case, the general expression for the relative
normal-velocities between the two initial states
$(v_{12}^x)_{_{2S}}$ can be calculated from
(\ref{g_vrel}) with the velocities behind the shock waves
$v^x_3$ and $v^x_{3'}$ being determined through the jump
condition (\ref{v_b_s}). Because $p_1$ is the smallest
value that the pressure at the contact discontinuity
$p_3$ can take, the limiting value for the two shock
waves branch $({\tilde v^x}_{12})_{_{2S}}$ can be
expressed as
\begin{equation}
\label{vtv12s}
(\widetilde{v}^x_{12})_{_{2S}}=\lim_{p_3\rightarrow
	p_1}(v^x_{12})_{_{2S}} \ .
\end{equation}
Evaluating the limit (\ref{vtv12s}) basically involves
calculating the limits of $v^x_{1,{\cal C}}$ and
$v^x_{2,{\cal C}}$ for $p_3$ tending to $p_1$. Both these
limits are straightforward to calculate and are
\begin{eqnarray}
\label{ulims_a}
&&\lim_{p_3\rightarrow p_1} v^x_{1,{\cal C}} = 0  \ ,
\\
\label{ulims_b}
&&\lim_{p_3\rightarrow p_1} v^x_{2,{\cal C}} =
	\frac{v^x_2-\bar{v}^x_{3'}} {1- v^x_2 \bar{v}^x_{3'}} \ ,
\end{eqnarray}
where $\bar{v}^x_{3'}$ is simply the value of $v^x_3$ for
$p_3=p_1$, i.e. 
\begin{equation}
\bar{v}^x_{3'}\equiv\lim_{p_3\rightarrow p_1} v^x_{3'} \ .
\end{equation}

	Using now the limits
(\ref{ulims_a})--(\ref{ulims_b}) and some lengthy but
straightforward algebra, the explicit analytic expression
for the limiting value of the two shock waves branch can
be calculated as
\begin{eqnarray}
\label{capo1}
(\widetilde{v}^x_{12})_{_{2S}}
	&=&-\lim_{p_3\rightarrow p_1}v^x_{2,{\cal C}}
	=\frac{(p_1-p_2)(1-v^x_2 \bar{V}_s)}
	{(\bar{V}_s-v^x_2)\{h_2 \rho_2
	(W_2)^2 [ 1-(v^x_2)^2 ] + p_1 - p_2\}}  \ .
\end{eqnarray}
Here $\bar{V}_s$ is the velocity of the shock wave
propagating towards the right in the limit of $p_3
\rightarrow p_1$ and an explicit expression for it can be
found in Appendix B in the case of an ideal fluid.
Expression (\ref{capo1}) will be discussed further
in Section~\ref{effect} but it sufficient to point out
here that the threshold value
$(\widetilde{v}^x_{12})_{_{2S}}$ does not depend on the
initial velocity in the state 1, $v_1$.

\subsection{$1\; {\cal R}_\leftarrow\; 
	3\; {\cal C}\; 3'\; {\cal S}_\rightarrow\; 2$: 
	One Shock and one Rarefaction Wave}
\label{case_ii}

	We next consider the wave-pattern consisting of a
rarefaction wave propagating towards the left and of a
shock wave propagating towards the right. Also in this
case, $(v_{12}^x)_{_{SR}}$ can be calculated from
(\ref{g_vrel}) with $v^x_{3'}$ being determined through
the jump condition (\ref{v_b_s}) and $v^x_3$ from the
numerical integration of equation (\ref{v_b_r}) in the
range $[p_1, p_{3}]$. Because $p_2$ is now the lowest
pressure in the unknown region behind the two nonlinear
waves, the limiting value for the one shock and one
rarefaction waves branch $({\tilde v^x}_{12})_{_{SR}}$
can be expressed as
\begin{equation}
(\tilde{v}^x_{12})_{_{SR}}=
	\lim_{p_3\rightarrow p_2}(v^x_{12})_{_{SR}} \ .
\end{equation}
In the limit $p_3 \rightarrow p_2$, the right-propagating
shock is suppressed, $v^x_{3'} \rightarrow v^x_{2}$ so
that
\begin{equation}
\lim_{p_3\rightarrow p_2} v^x_{2,{\cal C}} = 0 \ ,
\end{equation}
and
\begin{equation}
\label{capo2}
(\widetilde{v}^x_{12})_{_{SR}}=
	\lim_{p_3\rightarrow p_2}v^x_{1,{\cal C}} \ . 
\end{equation}
Defining now 
\begin{equation}
{\cal B}_1 \equiv \frac{1}{2}\log{ 
	\left( \frac{1+v_1^x}{1-v_1^x} \right) }\ ,
\end{equation}
and using (\ref{newraref}), it is readily obtained that
\begin{equation}
\label{capo2_analytic}
(\widetilde{v}^x_{12})_{_{SR}}=
	\lim_{p_3\rightarrow p_2} \tanh({\cal B}_1-{\cal B}) =
	\tanh\left(\int_{p_1}^{p_2} \frac{\sqrt{h^2 + 
	{\cal A}^2_1(1-c_s^2)}}
        {(h^2 + {\cal A}^2_1)\rho~c_s} dp \right)\ ,
\end{equation}
where the above integral can be evaluated numerically.  A
better look at the integral shows that only quantities in
the left state are involved (through the constant ${\cal
A}_1 \equiv h_1 W_1 v^t_1$) and that
$(\tilde{v}^x_{12})_{_{SR}}$ does not depend on the
initial velocity in the state 2, $v_2$. This property has
an important consequence that will be discussed in
Section~\ref{effect}.

\subsection{$1\; {\cal R}_\leftarrow\; 3\; 
	{\cal C}\; 3'\; {\cal R}_\rightarrow\; 2$: 
	Two Rarefaction Waves}
\label{case_iii}

	When the wave-pattern consists of two rarefaction
waves propagating in opposite directions,
$(v_{12}^x)_{_{2R}}$ can be calculated from
(\ref{g_vrel}) with the velocities behind the waves being
calculated using (\ref{newraref}) and (\ref{newraref2}).
Since the lowest value of the pressure behind the tails of the
rarefaction waves is zero, the limiting value for the two
rarefaction waves branch $({\tilde v^x}_{12})_{_{2R}}$ is
given by
\begin{equation}
\label{limit3}
(\tilde{v}^x_{12})_{_{2R}}=\lim_{p_3\rightarrow
	0}(v^x_{12})_{_{2R}} \ .
\end{equation}
Proceeding as in previous subsection, we can now express
$(\tilde{v}^x_{12})_{_{2R}}$ as
\begin{eqnarray}
\label{capo3}
(\tilde{v}^x_{12})_{_{2R}}&=&
	\frac{\bar{v}^x_{1,{\cal C}} - \bar{v}^x_{2,{\cal C}}}
	{1-  (\bar{v}^x_{1,{\cal C}}) (\bar{v}^x_{2,{\cal
	C}})} \ ,
\end{eqnarray}
where
\begin{eqnarray}
\label{capo3_a}	
\bar{v}^x_{1,{\cal C}} &=&
\tanh\left(\int_{p_1}^{0} \frac{\sqrt{h^2 + {\cal A}_1^2(1-c_s^2)}}
        {(h^2 + {\cal A}_1^2)\rho~c_s} dp \right) \ ,
	\\ \nonumber \\
\label{capo3_b}	
\bar{v}^x_{2,{\cal C}} &=&
\tanh\left(\int_{0}^{p_2} \frac{\sqrt{h^2 + {\cal A}_2^2(1-c_s^2)}}
        {(h^2 + {\cal A}_2^2)\rho~c_s} dp \right) \ ,
\end{eqnarray}
and where ${\cal A}_2 \equiv h_2 W_2 v^t_2$. While, the
determination of $(\tilde{v}^x_{12})_{_{2R}}$ requires
the numerical calculation of the integrals
(\ref{capo3_a}) and (\ref{capo3_b}), it has very little
practical importance as it marks the transition to a
wave-pattern with two rarefaction waves separated by
vacuum; this is a very rare physical configuration which
cannot be handled by a generic numerical code.
\begin{figure}[htb]
\centerline{
\psfig{file=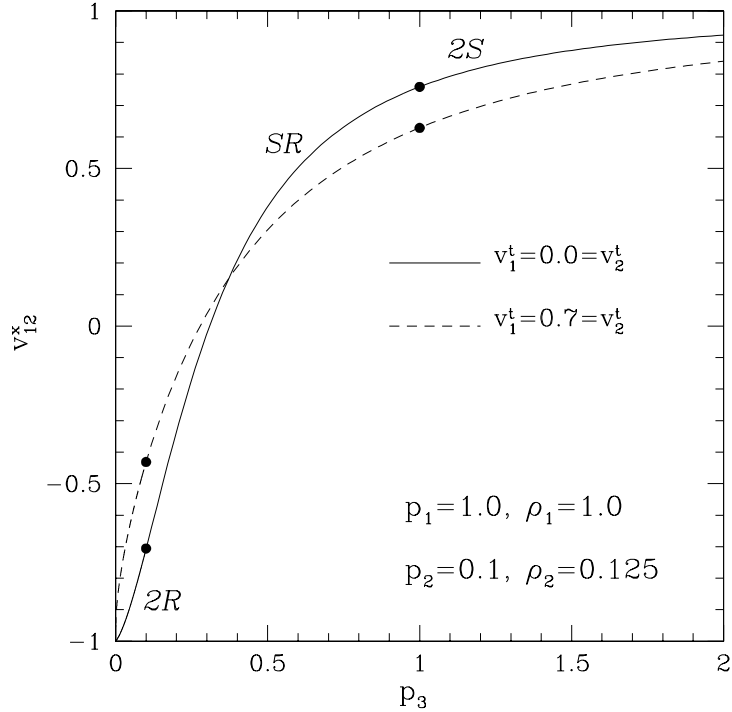,angle=0,width=10.0cm}
        }
\caption{
\label{fig0}
Relative normal-velocity between the two initial states
as a function of the pressure at the contact
discontinuity. Each curves is the continuous joining
(marked by the solid dots) of three different curves
corresponding respectively to two shock waves ($2S$), one
shock and one rarefaction wave ($SR$), and two
rarefaction waves ($2R$). The solid line refers to the
case of zero tangential velocities, while the dashed line
to the case in which $v^t_1=0.7=v^t_2$. The initial state
vectors are those of Sod's problem.}
\end{figure}

	Note that in computing (\ref{capo3}), both the
left state quantities and the right ones are involved
and, as a result, $(\tilde{v}^x_{12})_{_{2R}}$ will
depend on both $v_1$ and $v_2$. This property will be
important in the subsequent discussion in
Section~\ref{effect}.

	Fig.~\ref{fig0} shows the functional behaviour of
$v^x_{12}=v^x_{12}(p_3)$ and how this behaviour is
changed by the presence of nonzero tangential
velocities. The initial conditions are those of a
modified Sod's problem (Sod, 1978) in which $p_1=1.0$,
$\rho_1=1.0$, $v^x_1=0.0$, $p_2=0.1$, $\rho_2=0.125$,
$v^x_2=0.0$, and $\gamma=5/3$. Each of the two curves
shown is effectively the composition of three different
ones (joined at the solid dots) corresponding to
wave-patterns consisting of two shock waves ($2S$), one
shock and one rarefaction wave ($SR$), and two
rarefaction waves ($2R$). While the solid curve refers to
initial conditions with zero tangential velocities, the
dashed one is produced when nonzero tangential
velocities, $v^t_1= 0.7 =v^t_2$, are considered. Note
that also in this latter case, the three branches are
monotonically increasing with $p_3$ (a fundamental
property whose mathematical proof can be found in
Appendix A) but are all altered by the presence of
nonzero tangential velocities. The consequences of this
will be discussed in the next Section.

\section{Relativistic Effects}
\label{effect}

	The changes in the functional behaviour of
$v^x_{12}=v^x_{12}(p_3)$ introduced by nonzero initial
tangential velocities suggest that new qualitative
differences could be found in a relativistic
multidimensional Riemann problem. This was first
discussed in a recent paper (Rezzolla \& Zanotti 2002)
where the basic features of new relativistic effects were
briefly pointed out. This section is dedicated to a more
detailed discussion of how the changes in the functional
behaviour of $v^x_{12}=v^x_{12}(p_3)$ are responsible for
relativistic effects in the dynamics of nonlinear
waves. Before entering in the heart of the discussion,
however, it is useful to remind that in Newtonian
hydrodynamics a multidimensional Riemann problem does not
depend on the values of the tangential velocities in the
two initial states. Rather, different wave-patterns can
be produced only after a suitable change in either the
normal velocity, the rest-mass density or the
pressure. This is essentially due to the fact that
tangential velocities are not changed across Newtonian
nonlinear waves. In relativistic hydrodynamics, on the
other hand, this is not the case and is at the origin of
the effects discussed below.
\begin{figure}[htb]
\centerline{
\psfig{file=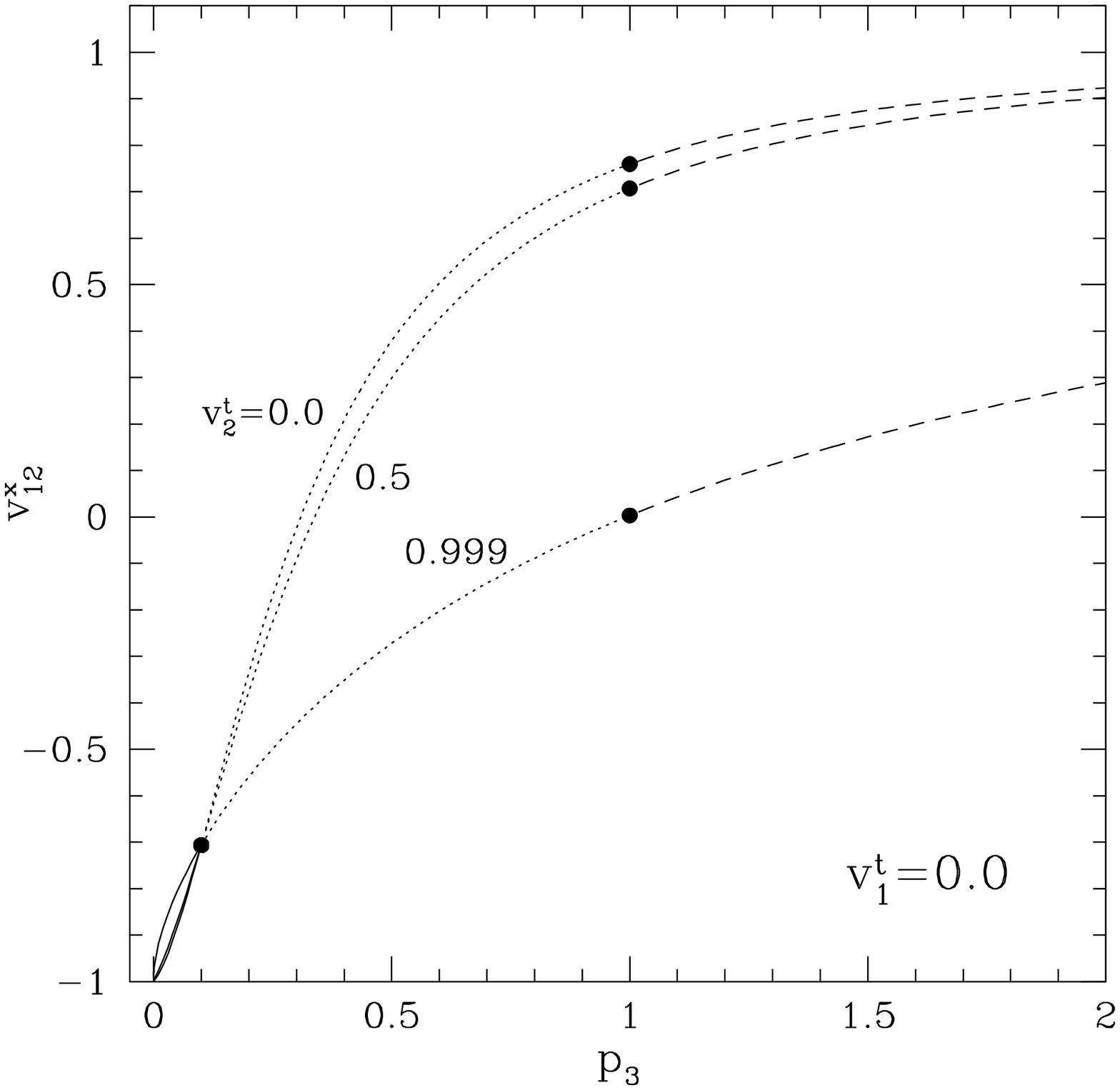,angle=0,height=7.0cm,width=7.0cm}
\hskip -1.0cm
\psfig{file=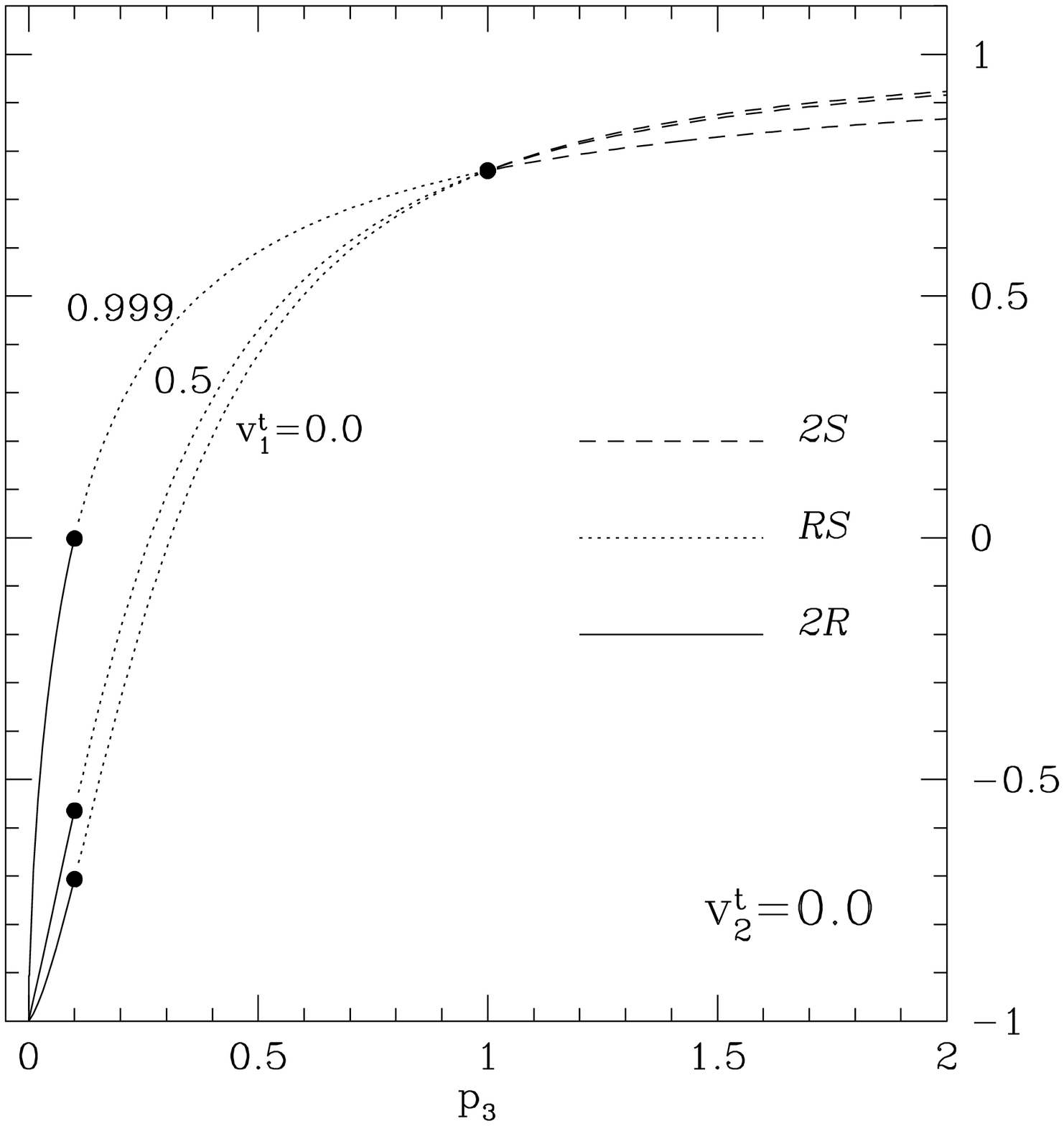,angle=0,height=7.0cm,width=7.0cm}
        }
\caption{
\label{fig1}
The same as in Fig.~\ref{fig0} but here the different
line types mark the different branches corresponding to
two shock waves (dashed line), one shock and one
rarefaction wave (dotted line), and two rarefaction waves
(continuous line), respectively. The two panels show how
the functional behaviour is modified when only one of the
initial tangential velocities is varied ($v^t_2$ for the
left panel and $v^t_1$ for the right one) while all the
other components of the initial state vectors are left
unchanged.}
\end{figure}

	Let us restrict our attention to a situation in
which the tangential velocity of only one of the two
initial states is varied. This is simpler than the
general case as it basically represents a one-dimensional
cross-section of the three-dimensional parameter space,
but it maintains all of the relevant properties. The two
panels of Fig.~\ref{fig1} show the relative
normal-velocity for the same initial conditions of
Fig.~\ref{fig0} where either $v^t_1$ or $v^t_2$ is varied
while all the other quantities of the initial state
vectors are left unchanged. Different line types mark the
different branches (joined at the filled dots) describing
the relative velocity corresponding to two shock waves
($2S$, dashed line), one shock and one rarefaction wave
($SR$, dotted line), and two rarefaction waves ($2R$,
continuous line), respectively.  Both panels of
Fig.~\ref{fig1} indicate that when tangential velocities
are present the relative normal-velocity is a function of
$p_3$ but also of $v^t_1$ and $v^t_2$\footnote{For the of
initial conditions chosen in Fig.~\ref{fig1} the position
of $({\tilde v}_{12})_{_{2R}}$ is very close to the limit
$-1$.}.

\begin{figure}[htb]
\centerline{
\psfig{file=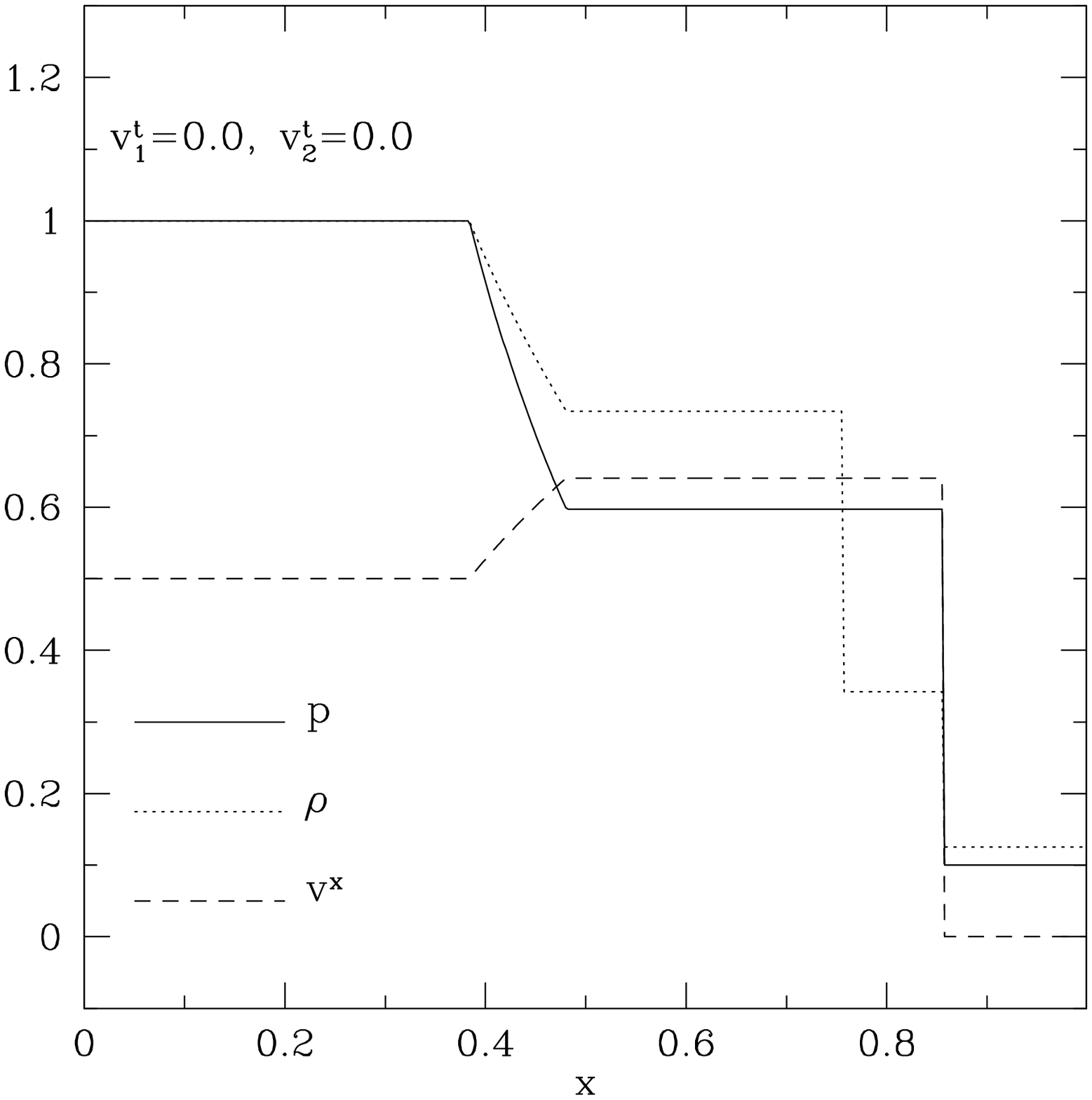,angle=0,height=7.0cm,width=7.0cm}
\hskip -1.0cm
\psfig{file=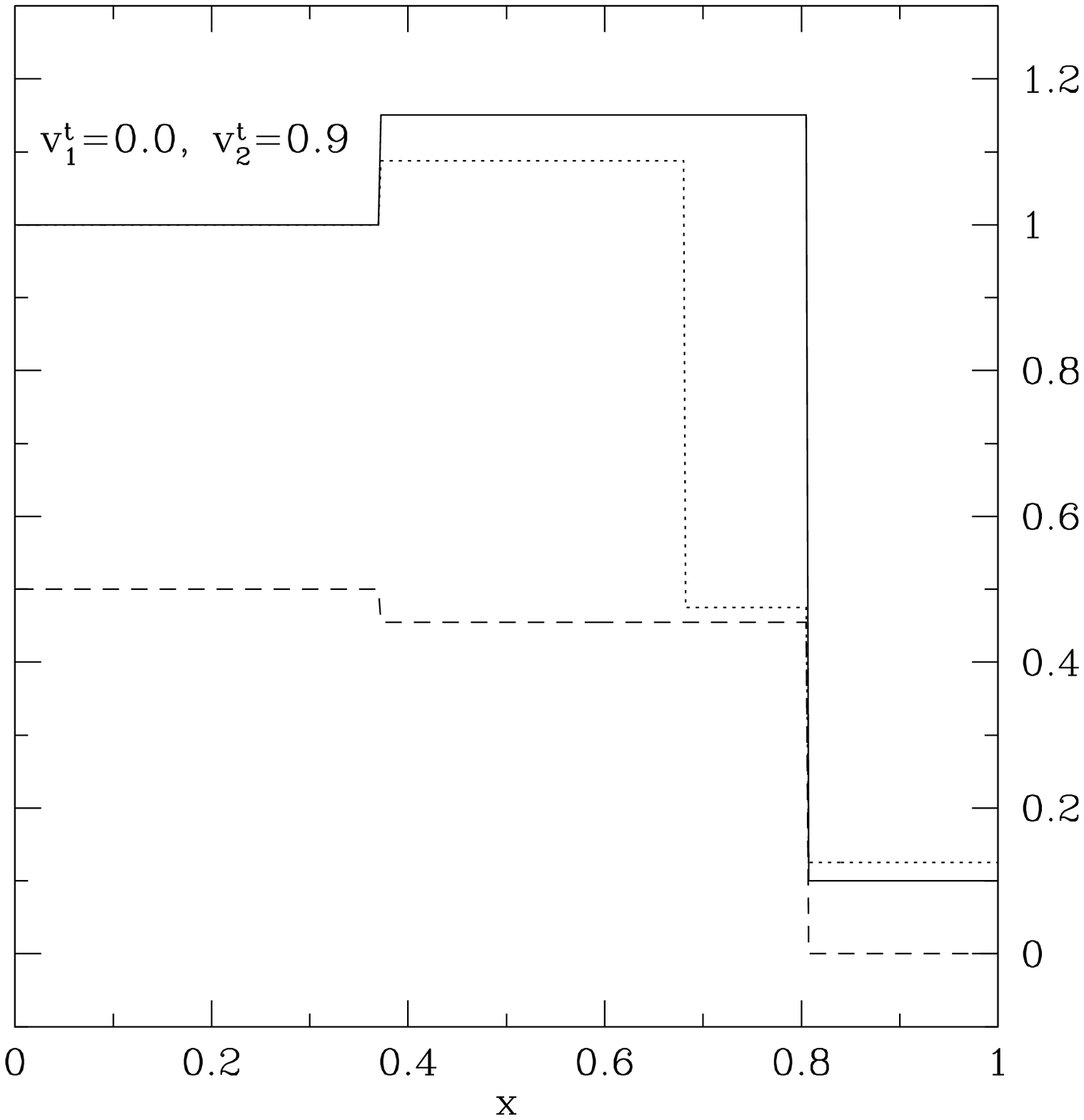,angle=0,height=7.0cm,width=7.0cm}
        }
\caption{Transition from a $SR$ wave-pattern to a $2S$
one. The left and right panels show the exact solution of
the Riemann problem corresponding to models (a) and (e)
in Table~I, respectively. The initial state vectors are
identical except for the values of $v^t_2$. Solid, dotted
and dashed profiles refer to pressure, rest-mass density
and normal velocity, respectively.
\label{fig3}
}
\end{figure}

	Consider, for instance, the case in which the
normal velocities are chosen to be $v^x_1 = 0.5, \; v^x_2
= 0.0$, and that there are no tangential velocities. In
this case, $(v^x_{12})_0 = 0.5$ and the left panel of
Fig.~\ref{fig1} shows that the solution to the Riemann
problem falls in the $SR$ branch, hence producing a
wave-pattern consisting of a shock and a rarefaction wave
moving in opposite directions. This is shown in more
detail in the left panel of Fig.~\ref{fig3} where the
different types of line show the solution of the Riemann
problem at a time $t>0$ for the pressure (continuous
line), the rest-mass density (dotted line) and the
velocity (dashed line).

	However, if we now maintain the {\it same}
initial conditions but allow for nonzero tangential
velocities in state 2, the left panel of Fig.~\ref{fig1}
also shows that the solution to the Riemann problem can
fall in the $2S$ branch, hence producing a wave-pattern
consisting of two shock waves moving in opposite
directions. This is shown in the right panel of
Fig.~\ref{fig3} which illustrates the solution of the
same Riemann problem but with initial tangential
velocities $v^t_1=0.0$ and $v^t_2=0.9$. Note that except
for the tangential velocities, the solutions in
Figs.~\ref{fig3} have the same initial state-vectors but
different intermediate ones (i.e.
$p_3,\;\rho_3,\;\rho_{3'}$, and $v^x_3$).

	The Riemann problem shown in Fig.~\ref{fig3} is
only one possible example but shows that a change in the
tangential velocities can produce a {\em smooth
transition from one wave-pattern to another while
maintaining the initial states unmodified}. Furthermore,
because the coupling among the different states is
produced by the Lorentz factors this is not sensitive on
the sign chosen for the tangential velocity.
\begin{figure}[htb]
\centerline{
\psfig{file=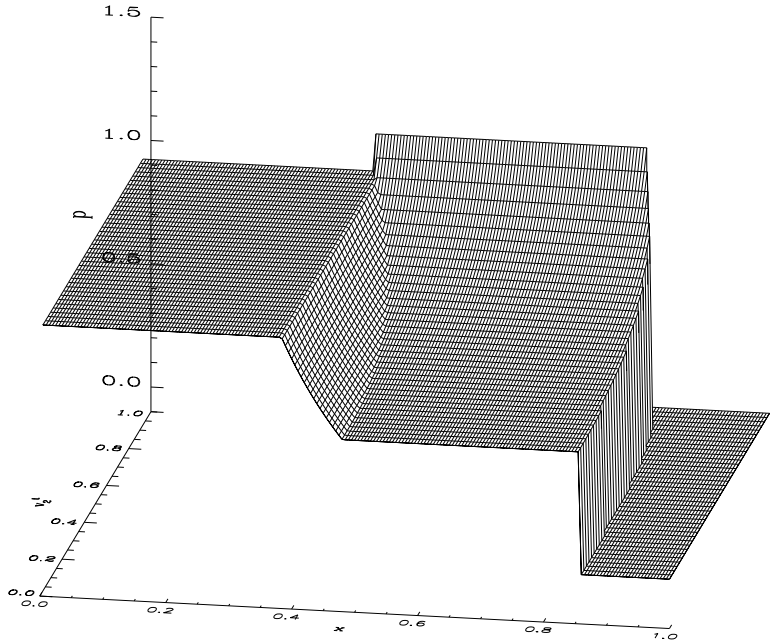,angle=90,width=12.0cm} }
\caption{Sequence of solutions for the pressure in Sod's
problem. The initial tangential velocity $v^t_2$ is
gradually increased from $0$ to $0.9$.  The first and
last solution of this sequence are also plotted in
Fig.~\ref{fig3}
\label{fig4}
}
\end{figure}
The transition from one wave-pattern to the other is
better illustrated in Fig.~\ref{fig4} where we have
collected in a three-dimensional plot a sequence of
solutions for the pressure in which $v^t_2$ is gradually
increased from $0$ to $0.9$. Note that when $v^t_2=0$,
the $SR$ wave-pattern is well defined and the pressure at
the contact discontinuity is intermediate between $p_1$
and $p_2$. Note also that as $v^t_2$ is increased, the
wave-pattern gradually changes, $p_3$ increases up until
it becomes larger than $p_1$, signalling the transition
to a $2S$ wave-pattern.

	Interestingly, the transition does not need to
always produce a solution consisting of two shock
waves. Suppose, in fact, that the normal velocities are
now chosen to be $v^x_1=0.0,\;v^x_2=0.5$. We can repeat
the considerations made above and start by examining the
wave-pattern produced when there are zero tangential
velocities.
\begin{figure}[htb]
\centerline{
\psfig{file=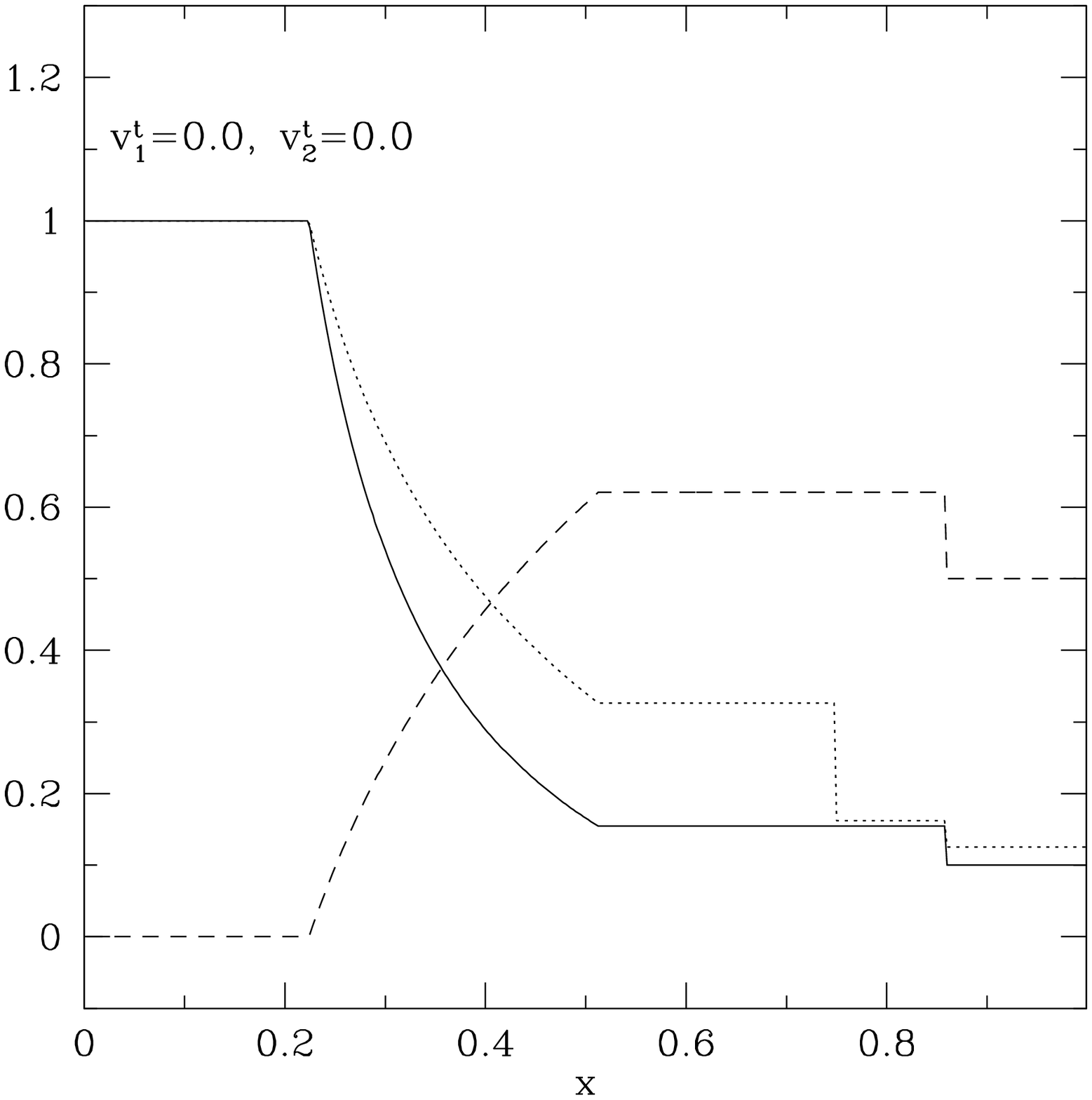,angle=0,height=7.0cm,width=7.0cm}
\hskip -1.0cm
\psfig{file=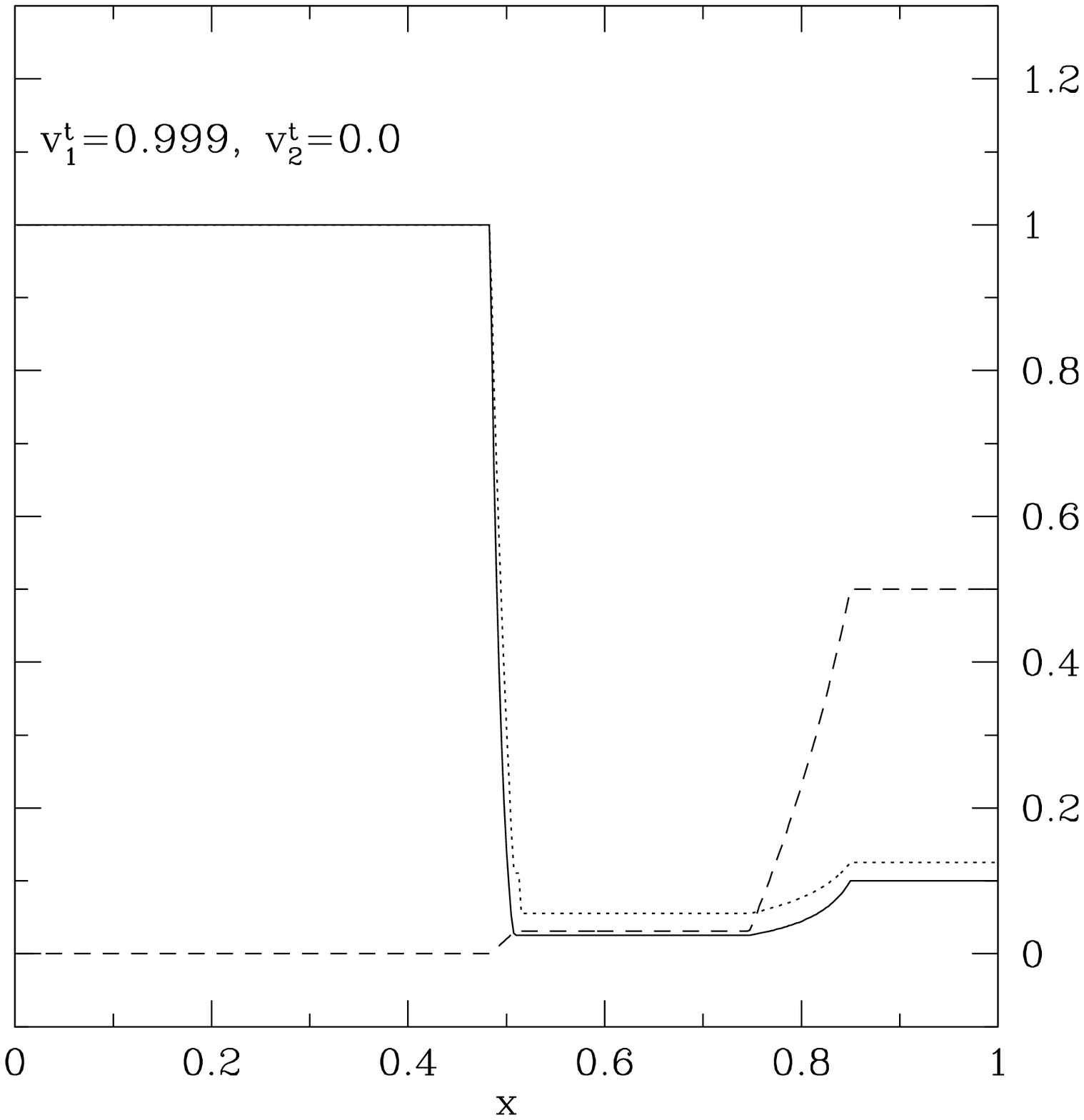,angle=0,height=7.0cm,width=7.0cm}
        }
\caption{The same as in Fig.~\ref{fig3} but for models
(h) and (n) in Table~I. Also in this case the initial
state vectors are identical except for the values of
$v^t_1$. Note that in the right panel the
left-propagating rarefaction wave covers a very small
region of the flow and is closely followed by the contact
discontinuity.
\label{fig5}
}
\end{figure}
In this new setup, $(v^x_{12})_0 = -0.5$ and the right
panel of Fig.~\ref{fig1} shows that the solution to the
Riemann problem still falls in the $SR$ branch
(cf. dashed line), with the corresponding solution at a
time $t>0$ being presented in the left panel of
Fig.~\ref{fig5} (Note that the wave-patterns in
Fig.~\ref{fig3} and~\ref{fig5} both consist of a shock
and a rarefaction wave, but have alternating initial
normal velocities.).
\begin{figure}[htb]
\centerline{
\psfig{file=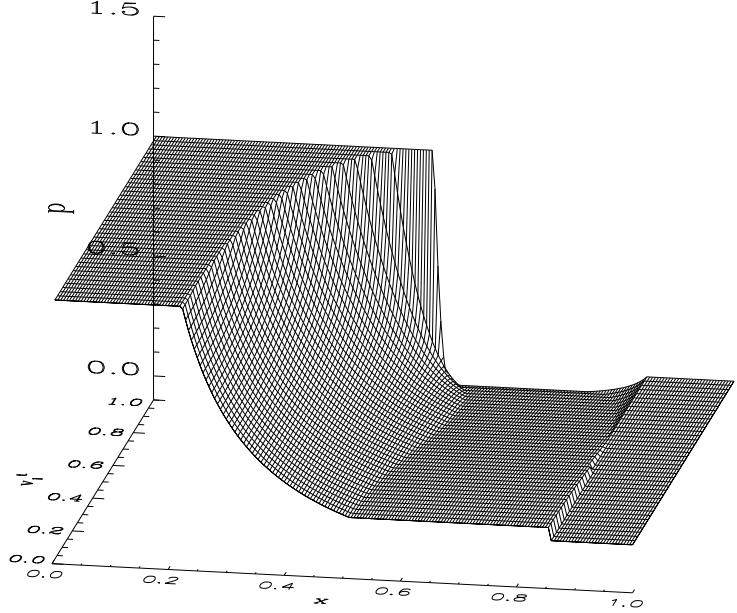,angle=0,width=12.0cm}
        }
\caption{The same as in Fig.~\ref{fig4} but here the
initial tangential velocity $v^t_1$ is gradually
increased from $0$ to $0.999$. The first and last
solution of this sequence are also plotted in
Fig.~\ref{fig5}
\label{fig6}
}
\end{figure}

	When nonzero tangential velocities are now
considered in state 1, the right panel of Fig.\ref{fig1}
shows that $(v^x_{12})_0$ can fall in the $2R$ branch,
hence producing a wave-pattern consisting of two
rarefaction waves moving in opposite directions. The
solution to this Riemann problem is shown in right panel
of Fig.~\ref{fig5} where we have chosen initial
tangential velocities $v^t_1=0.999$ and $v^t_2=0.0$. In
this case too, it should be noted that, except for the
tangential velocities, the solutions in the two panels of
Fig.~\ref{fig5} have the same initial state-vectors but
different intermediate ones.

	In analogy with Fig.~\ref{fig4}, we have
collected in Fig.~\ref{fig6} a sequence of solutions for
the pressure in which $v^t_1$ is gradually increased from
$0$ to $0.999$. Here too, when $v^t_2=0$, the $SR$
wave-pattern is well defined and the pressure at the
contact discontinuity is intermediate between $p_1$ and
$p_2$.
\begin{table}
\begin{center}
\label{table1}
\begin{tabular}{|c|cccc|cccc|c|}
\cline{1-10} &&&&&&&&& \\
Model 	& $\quad v^x_1  \quad$ & $\quad v^x_2 	  \quad$
      	& $\quad v^t_1  \quad$ & $\quad v^t_2 	  \quad$ 
	& $\quad p_*    \quad$ & $\quad v^x_* 	  \quad$ 
	& $\quad \rho_3 \quad$ & $\quad \rho_{3'} \quad$ 
	& wave-pattern \\
\cline{1-10} &&&&&&&&& \\
(a) & 0.5 & 0.0 & 0.0 & 0.000 & 0.597 & 0.640 & 0.734 & 0.342 & SR\\
(b) & 0.5 & 0.0 & 0.0 & 0.300 & 0.621 & 0.631 & 0.751 & 0.349 & SR\\
(c) & 0.5 & 0.0 & 0.0 & 0.500 & 0.673 & 0.611 & 0.788 & 0.364 & SR\\
(d) & 0.5 & 0.0 & 0.0 & 0.700 & 0.787 & 0.570 & 0.866 & 0.394 & SR\\
(e) & 0.5 & 0.0 & 0.0 & 0.900 & 1.150 & 0.455 & 1.088 & 0.474 & 2S\\ 
(f) & 0.5 & 0.0 & 0.0 & 0.990 & 2.199 & 0.212 & 1.593 & 0.647 & 2S\\
(g) & 0.5 & 0.0 & 0.0 & 0.999 & 3.011 & 0.078 & 1.905 & 0.750 & 2S\\
\cline{1-10} &&&&&&&&& \\
(h) & 0.0 & 0.5 & 0.000 & 0.0 & 0.154 & 0.620 & 0.326 & 0.162 & SR\\ 
(i) & 0.0 & 0.5 & 0.300 & 0.0 & 0.139 & 0.594 & 0.306 & 0.152 & SR\\ 
(j) & 0.0 & 0.5 & 0.500 & 0.0 & 0.115 & 0.542 & 0.274 & 0.136 & SR\\ 
(k) & 0.0 & 0.5 & 0.700 & 0.0 & 0.085 & 0.450 & 0.228 & 0.113 & 2R\\
(l) & 0.0 & 0.5 & 0.900 & 0.0 & 0.051 & 0.280 & 0.168 & 0.084 & 2R\\
(m) & 0.0 & 0.5 & 0.990 & 0.0 & 0.031 & 0.095 & 0.123 & 0.061 & 2R\\
(n) & 0.0 & 0.5 & 0.999 & 0.0 & 0.026 & 0.031 & 0.110 & 0.052 & 2R\\
\cline{1-10} 
\end{tabular}
\end{center}
\smallskip
\caption{Solution of the modified Sod's problem at
$t=0.4$. All models refer to an ideal EOS with
$\gamma=5/3$ and share the same values of pressure and
rest-mass density: $p_1=1.0$, $\rho_1=1.0$, $p_2=0.1$,
$\rho_2=0.125$. The only differences present in the
problems considered are in the normal relative velocity
and in the tangential velocities. These quantities are
reported in the first three columns, while the remaining
ones show a few relevant quantities of the solution in
the newly formed region as well as the wave pattern
produced.}
\end{table}
However, as $v^t_1$ is increased, the wave-pattern
gradually changes, $p_3$ decreases until it becomes
smaller than $p_2$, signalling the transition to a $2R$
wave-pattern. Note that while this happens, the region of
the flow covered by the rarefaction wave becomes
progressively smaller.

	In Table~I we have summarized a few of the
solutions shown in Figs.~\ref{fig4} and~\ref{fig6},
presenting numerical values for all of the relevant
quantities in the Riemann problem when different
combinations of the tangential velocities are used.

\begin{figure}[t]
\centerline{
\psfig{file=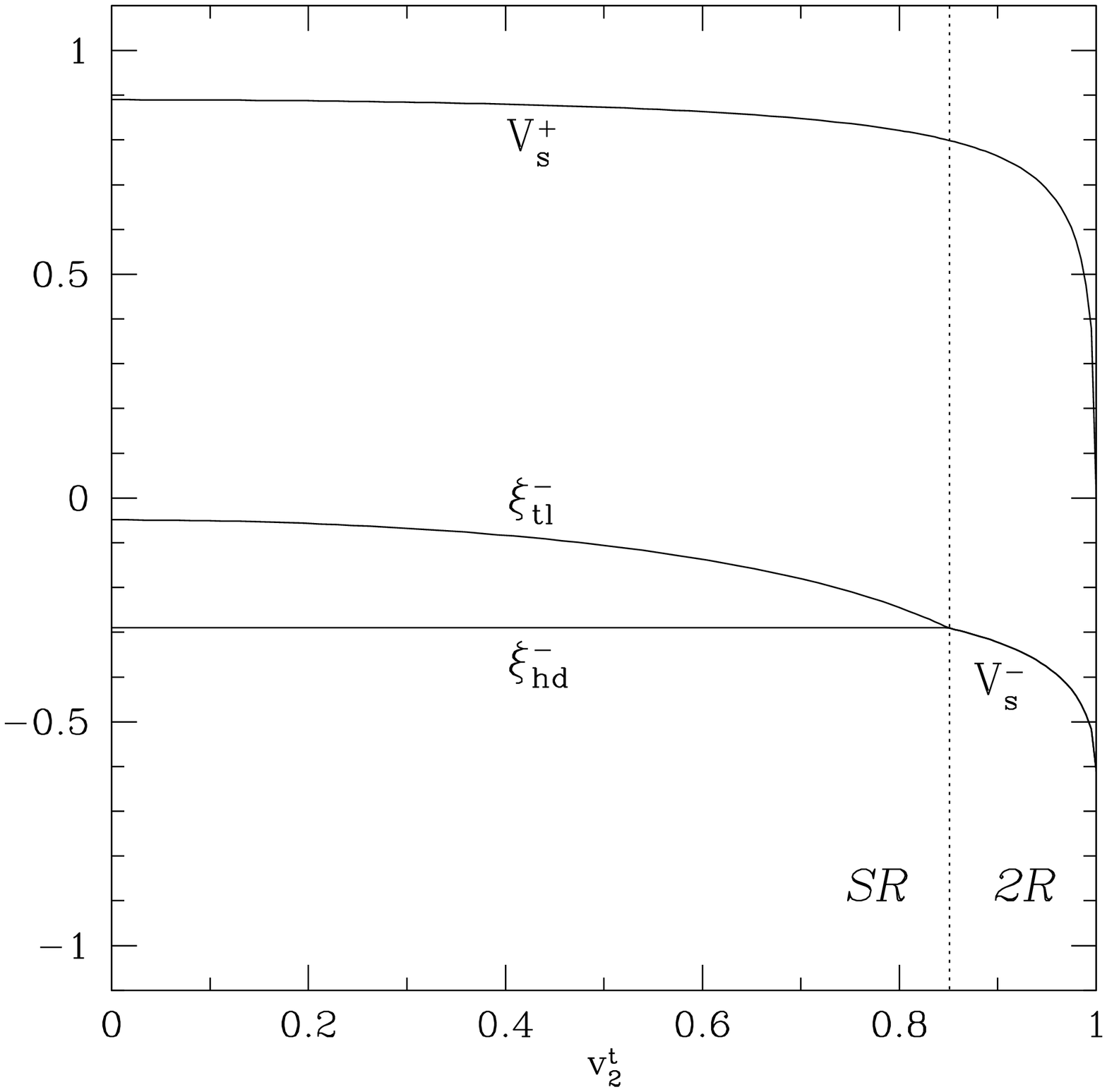,angle=0,height=7.0cm,width=7.0cm}
\hskip -1.0cm
\psfig{file=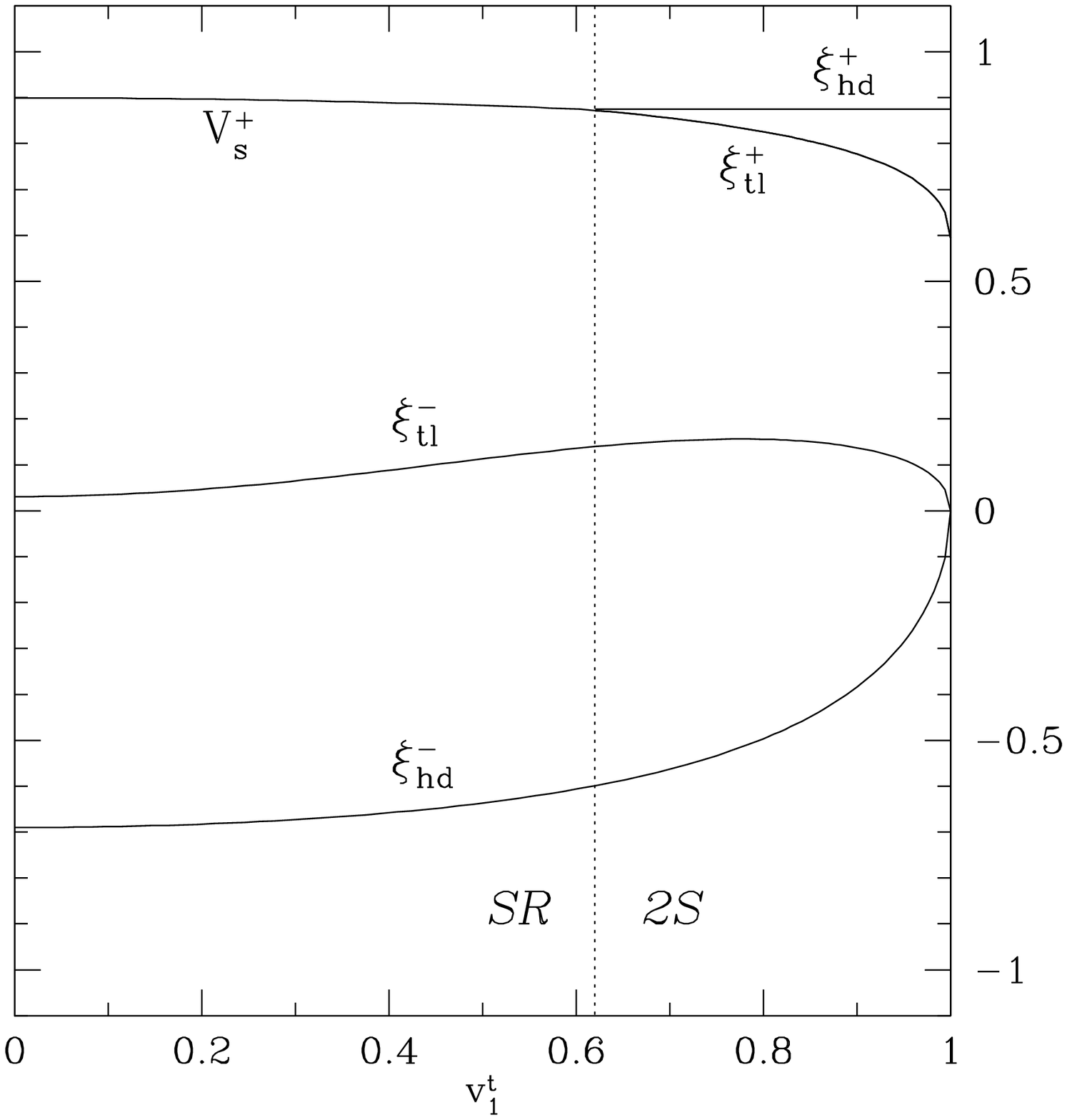,angle=0,height=7.0cm,width=7.0cm}
        }
\caption{Velocities of the various nonlinear waves when
the tangential velocities $v^t_1$ and $v^t_2$ are varied
separately. The initial conditions are those of Sod's
problem and the different curves refer to the head and
tail of a left-propagating rarefaction wave
(i.e. $\xi^{-}_{\rm tl}$, $\xi^{-}_{\rm hd}$), to the
head and tail of a right-propagating rarefaction wave
(i.e. $\xi^{+}_{\rm tl}$, $\xi^{+}_{\rm hd}$), and to a
left or a right-propagating shock wave (i.e. $V^{-}_{s}$,
$V^{+}_{s}$).
\label{fig7}}
\end{figure}

	To gain a better insight in these effects it can
be instructive to consider how the velocities at which
the various nonlinear waves propagate in the unperturbed
media change when the tangential velocities $v^t_1$ and
$v^t_2$ are varied separately. This information is
contained in Fig.~\ref{fig7} in which different curves
show the behaviour of the head and tail of a
left-propagating rarefaction wave (i.e. $\xi^{-}_{\rm
tl}$, $\xi^{-}_{\rm hd}$), of the head and tail of a
right-propagating rarefaction wave (i.e. $\xi^{+}_{\rm
tl}$, $\xi^{+}_{\rm hd}$), and of a left or a
right-propagating shock wave (i.e. $V^{-}_{s}$,
$V^{+}_{s}$). The left panel of Fig.~\ref{fig7}, in
particular, shows the transition from a $SR$ to a $2S$
wave-pattern with the dotted line marking the value of
$v^t_2$ at which this occurs. Similarly, the right panel
shows the transition from a $SR$ to a $2R$ wave-pattern
and the dotted line is again used to mark the value of
$v^t_1$ distinguishing the two regions of the parameter
space. A number of interesting features can be noted and
some of these were pointed out also by Pons {\it et al}
(2000). Firstly, the speed of the head of a rarefaction
wave propagating towards a region of constant tangential
velocity is constant, or, stated it differently,
$\xi^{-}_{\rm hd}$ does not depend on $v^t_2$.  Secondly,
the velocity of the waves converges to zero if they
propagate in regions with increasingly large tangential
velocities. Thirdly, the values of $v^t$ at which the
speeds of the head and tail of the rarefaction wave
coincide, mark the transition from one wave-pattern to
another and are indicated with vertical dotted lines in
Fig.~\ref{fig7}.

	As mentioned before, the appearance of these new
relativistic effects is related to the behaviour of the
function $v^x_{12}=v^x_{12}(p_3)$ for different values of
the initial tangential velocities and in particular to
how the three branches composing the curve change under
variation of $v^t_{1,2}$. As a result, the occurrence of
these effects can be recast into the study of the
dependence of $({\tilde v^x}_{12})_{_{2S}}, ({\tilde
v^x}_{12})_{_{SR}}$ and $({\tilde v^x}_{12})_{_{2R}}$ on
the tangential velocities. Using expressions
(\ref{capo1}), (\ref{capo2_analytic}), and (\ref{capo3}),
this dependence can be summarised as follows
\begin{equation}
\label{dep}
({\tilde v^x}_{12})_{_{2S}} = 
	({\tilde v^x}_{12})_{_{2S}}(v^t_2) \ ,
\hskip 0.5cm
({\tilde v^x}_{12})_{_{SR}} = 
	({\tilde v^x}_{12})_{_{SR}}(v^t_1)\ ,
\hskip 0.5cm
({\tilde v^x}_{12})_{_{2R}} = 
	({\tilde v^x}_{12})_{_{2R}}(v^t_1, v^t_2) \ ,
\end{equation}
and can be best studied by considering the limits of
$({\tilde v^x}_{12})_{_{2S}}, ({\tilde
v^x}_{12})_{_{SR}}$ and $({\tilde v^x}_{12})_{_{2R}}$
when $W_{1,2} \rightarrow \infty$. In the case of a $2S$
wave-pattern, expression (\ref{capo1}) simply indicates
that
\begin{equation}
\label{cond_a}
\lim_{W_2\rightarrow\infty}({\tilde v^x}_{12})_{_{2S}}=0
	\ .
\end{equation}
This result is also shown in the left panel of
Fig.~\ref{fig1}, where the right solid dot converges to
zero as $W_2 \rightarrow \infty$, while the left one does
not vary. The limit (\ref{cond_a}) can also be used to
deduce that for any $(v^x_{12})_0>0$, there exists a
value ${\bar W_2}$ of $W_2$ such that
\begin{equation}
\label{cond_ab}
(v^x_{12})_0 > ({\tilde v}^x_{12})_{_{2S}}\hskip 2.0cm 
	{\rm for}\quad W_2 > \bar{W_2}
	\ .
\end{equation}
A direct consequence of (\ref{cond_ab}) is that given a
Riemann problem having initial state vectors with
positive relative normal-velocity and producing a $SR$
wave-pattern, it is always possible to transform it into
a $2S$ wave-pattern by increasing the value of the
initial tangential velocity in the state of initial lower
pressure.

	In the case of a $SR$ wave-pattern we refer to
(\ref{capo2_analytic}) to notice that in the limit of
$W_1 \rightarrow \infty$ the integrand vanishes (${\cal
A}_1 \rightarrow \infty$) and therefore:
\begin{equation}
\label{cond_b}
\lim_{W_1\rightarrow\infty}({\tilde v^x}_{12})_{_{SR}}=0
\ .
\end{equation}
As for the previous one, the limit (\ref{cond_b}) can be
deduced from the right panel of Fig.~\ref{fig1}, where
the left solid dot converges to zero as $W_1 \rightarrow
\infty$, while the right one does not vary. Also in this
case the limit (\ref{cond_b}) can be used to conclude
that for any $(v^x_{12})_0 < 0$, there exists a value
$\bar{W_1}$ of $W_1$ such that
\begin{equation}
\label{cond_bb}
(v^x_{12})_0 < ({\tilde v}^x_{12})_{_{SR}} \hskip 2.0cm 
	{\rm for}\quad W_1 > \bar{W_1}
	\ ,
\end{equation}
and therefore causing an initial $SR$ wave-pattern
solution to become a $2R$ one as a consequence of an
increased tangential velocity in the state of initial
higher pressure\footnote{Note that it is not possible to
provide analytic expressions for $\bar{W_2}$ nor for
$\bar{W_1}$.}.

	Overall, expressions (\ref{cond_a}) and
(\ref{cond_b}) indicate that for tangential velocities
assuming increasingly larger values, the $SR$ branch of
the $v^x_{12}$ curve spans a progressively smaller
interval of relative normal-velocities. When the
tangential velocities reach their asymptotic values, the
$SR$ branch reduces to a point. In practice, therefore,
the main effect introduced by relativistic tangential
velocities in a Riemann problem is that of disfavouring
the occurrence of a wave-pattern consisting of a shock
and a rarefaction wave.

	For completeness we also report the limit of the
relative normal-velocity marking the branch of two
rarefaction waves separated by vacuum. In this case, the
limit is taken for both $W_1$ and $W_2$ tending to
infinity, and using (\ref{capo3_a})-(\ref{capo3_b})
yields
\begin{equation}
\label{cond_c}
\lim_{W_{1,2}\rightarrow\infty}({\tilde
	v^x}_{12})_{_{2R}}=0 \ .
\end{equation}

	Because the effects discussed in this Section
have a purely special relativistic origin they might
conflict with our physical intuition, especially when the
latter is based on the knowledge of the Riemann problem
in Newtonian hydrodynamics. However, the behaviour
reported here is typical of those special relativistic
phenomena involving Lorentz factors including also
tangential velocities. A useful example in this respect
is offered by the relativistic transverse-Doppler effect,
in which the wavelength of a photon received from a
source moving at relativistic speeds changes also if the
source has a velocity component orthogonal to the
direction of emission of the photon (Rindler, 1980). In
this case too, a Lorentz factor including the transverse
velocity is responsible for the effect.

	Finally, it should be pointed out that there
exists a set of initial conditions for which these new
relativistic effects cannot occur. These initial
conditions are those in which $v^x_1=v^x_2$ as in the
classic ``shock-tube'' problem, where $v^x_1=0=v^x_2$. In
this case, in fact, $(v^x_{12})_0 = 0$ and, because of
the limits (\ref{cond_a}) and (\ref{cond_b}), the
solution of the Riemann problem will be given by a
wave-pattern consisting of a shock and a rarefaction
wave, independently of the values of the tangential
velocities.

\section{Conclusions}

	We have shown that and efficient solution of the
exact Riemann problem in multidimensional relativistic
flows can be obtained after exploiting the properties of
the invariant expression for the relative normal-velocity
between the two initial states. The new procedure
proposed here is the natural extension of a similar
method presented for the exact solution of the Riemann
problem in one-dimensional relativistic flows (Rezzolla
\& Zanotti, 2001). Using information contained in the
initial state vectors, this approach predicts the
wave-pattern that will be produced in the Riemann
problem, determines the set of equations to be solved and
brackets the interval in pressure where the solution
should be sought. Because logically straightforward, this
approach results into an algorithm which is very easy to
implement numerically and work is now in progress to
assess the computational speedup in multidimensional
codes.

	An important aspect of this strategy is that it
naturally points out relativistic effects that can take
place whenever the initial relative velocity normal to
the initial surface of discontinuity is nonzero. When
this is the case, in fact, the tangential velocities can
affect the solution of the Riemann problem and cause a
transition from one wave-pattern to another one. More
specifically, by varying the tangential velocities on
either side of the initial discontinuity while keeping
the remaining state vectors unchanged, the nonlinear
waves involved in the solution Riemann problem can change
from rarefaction waves to shock waves and
vice-versa. These effects have a purely relativistic
nature, do not have a Newtonian counterpart and could be
relevant in several astrophysical scenarios, such as
those involving relativistic jets or $\gamma-$ray bursts,
in which nonlinear hydrodynamical waves with large
Lorentz factors and complex multidimensional flows are
expected (Blandford 2002, Meszaros 2002). 

	As a final remark it is worth pointing out that
while the content of this paper is focussed on special
relativistic hydrodynamics and flat spacetimes, the local
Lorentz invariance allows to extend the results discussed
here also to curved spacetimes and general relativistic
numerical calculations (Pons {\it et al} 1998, and Font
2000 for a review).

\bigskip
\bigskip
\appendix{\large \bf Appendix A: Monotonicity of the relative 
	velocity as function of $p_*$}
\bigskip

	This Appendix is devoted to the proof that
$v^x_{12}$ is a monotonic function of $p_*$; as mentioned
in the main text, this is an important property and the
basis of our approach. 

	With our choice of considering the initial left
state as the one with highest pressure, the proof of
monotonicity will be obtained if we show that $v^x_{12}$
is a monotonically {\it increasing} function of
$p_*$. Indicating then with a $'$ the first derivative
with respect to $p_*$ and dropping the upper index $x$ in
the notation for the normal velocities, it is
straightforward to derive that the first derivative of
expression (\ref{g_vrel}) is given by
\begin{equation}
\label{d_vrel}
v'_{12}= \frac{v'_{1,{\cal C}}
(1-v^2_{2,{\cal C}}) - v'_{2,{\cal C}}(1-v^2_{1,{\cal C}})}
{(1-v_{1,{\cal C}}\;v_{2,{\cal C}})^2} \ .
\end{equation}	
A rapid look at expression (\ref{d_vrel}) suggests that
the proof that $v^x_{12}$ is monotonically increasing
will follow if it can be shown that $v'_{1,{\cal C}}$ and
$-v'_{2,{\cal C}}$ are both positive. On the other hand,
using equations (\ref{1c}) and (\ref{2c}), the proof of
the monotonicity will follow from showing that $(v^x_3)'
< 0 $ and $(v^x_{3'})' > 0$. While these inequalities
must hold irrespectively of the nonlinear wave
considered, the proofs will be different for the
different waves considered.
	
 	In the case when a rarefaction wave is present,
the proof is indeed straightforward. According to
(\ref{v_b_r}) and (\ref{o_m_xiv}), in fact, $(v^x)'$
across the rarefaction wave is negative when the
rarefaction wave propagates towards the left [implying
that $(v^x_3)'$ is negative] and it is positive when the
rarefaction wave propagates towards the right [implying
that $(v^x_{3'})'$ is positive].

\begin{figure}[t]
\centerline{
\psfig{file=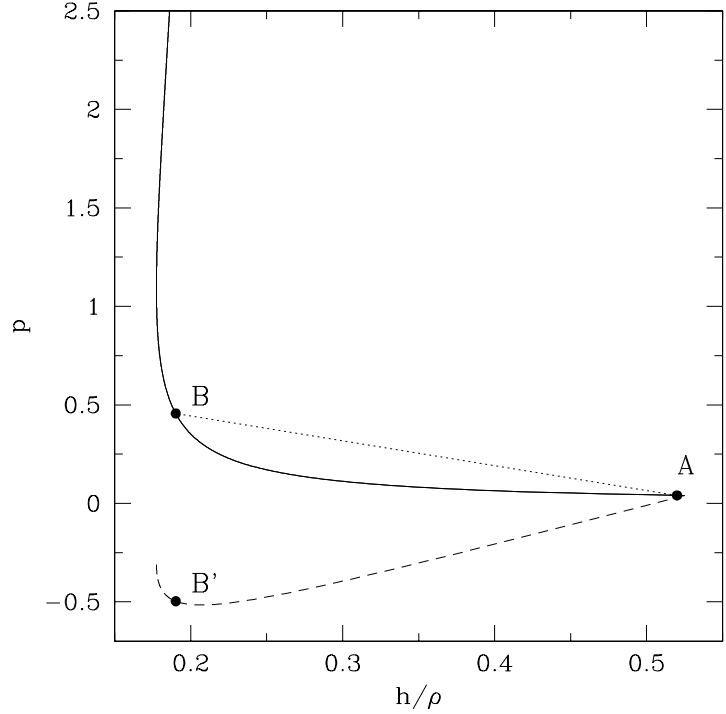,angle=0,width=10.0cm}
        }
\caption{The solid curve represents the Taub adiabat and
the points A and B on it the states ahead and behind a
left-propagating shock wave. See the main text for a
discussion. \label{fig8} }
\end{figure}

	If a shock wave is present, on the other hand, a
proof for the most general case and in terms of simple
algebraic relations and cannot be given. On the other
hand, a rather simple analytic proof can be found in the
simpler case in which $v^x_1=v^x_2=0$; while this is
certainly not the most general case, numerical
calculations have shown that the result holds in general.
Consider therefore a shock wave propagating towards the
left\footnote{A similar analysis can be repeated for the
right-propagating shock wave}; after lengthy but
straightforward calculations it is possible to show that
\begin{equation}
\label{cond1}
(v^x_3)' = \frac{H_1 (V_s - v^x_1)(1-V_s
v^x_1) - H_1 \Delta p (1 - (v^x_1)^2)V'_s - V'_s (\Delta
p)^2}{[H_1 (V_s - v^x_1) + \Delta p V_s]^2} \ ,
\end{equation}
where we have set $H_1 \equiv h_1\rho_1 W_1^2$ and
$\Delta p \equiv p-p_1>0$, and where $p$ is the pressure
behind the shock. If we now impose that $v^x_1=0$, we can
write the derivative of (\ref{velshock}) as
\begin{equation}
\label{vs'}
\frac{V'_s}{V_s}=\frac{J'}{J}\frac{\rho_1^2
W_1^2}{J^2+\rho_1^2 W_1^2 } \ .
\end{equation}
Inserting (\ref{vs'}) in (\ref{cond1}) we can conclude
that $(v^x_3)'$ is negative if and only if
\begin{equation}
\label{cond2}
J(\rho_1^2 W_1^2+J^2)h_1 - \rho_1 \Delta p J'(\rho_1 h_1
	W_1^2 + \Delta p)<0 \ .
\end{equation}
Using (\ref{flux2}) to calculate $J'$, one finds that
(\ref{cond2}) can be written as
\begin{equation}
\label{cond3}
-\frac{\rho_1(H_1 - \Delta p)}{J^2} -
	2 h_1 <
	\frac{1}{\gamma(\gamma-1)}\frac{\rho_1(H_1 + \Delta p)}{\rho^2}
	\left[\frac{1}{\epsilon} -
	\gamma(\gamma-2)\right] \ ,
\end{equation}
where $\rho$ and $\epsilon$ are the rest-mass density and
the specific internal energy behind the shock
front. Because the right hand side of (\ref{cond3}) is
always positive for any $\gamma\leq 2$, the condition for
monotonicity (\ref{cond3}) will be satisfied if its left
hand side is negative, i.e. if
\begin{equation}
\label{cond4}
-J^2 <- \rho_1\frac{\Delta p - H_1}{2h_1} \equiv -\alpha \ .
\end{equation}
At this point the proof can be continued graphically and
making use of the Taub adiabat. In the plane
$(h/\rho,p)$, in fact, the Taub adiabat (\ref{Taub2})
selects the points solutions of the hydrodynamical
equations across a shock wave, therefore connecting the
state ahead of the shock front with the one behind it. In
Fig.~\ref{fig8}, this curve is indicated with a solid
line and we have indicated with the points A and B the
states ahead (region 1) and behind (region 3) the shock
front. Once an initial state A has been chosen, the mass
flux will determine the point B of the Taub adiabat
solution of the Rankine-Hugoniot relations. Because of
this, the slope of the chord connecting the point A with
B (shown as a dotted line in Fig.~\ref{fig8}) is equal to
$-J^2$. Indicated with a dashed line, Fig.~\ref{fig8}
also shows the equivalent of the Taub adiabat passing
through the state A but having mass flux equal to
$\alpha^{1/2}$, i.e.
\begin{equation}
p = -\alpha\left(\frac{h}{\rho}-\frac{h_1}{\rho_1}\right)
	+ p_1 \ .
\end{equation}
The point B' on such a curve represents the state behind
the shock wave and, as it is clearly shown in
Fig.~\ref{fig8}, the slope of the chord AB' is always
larger than the corresponding slope for the chord AB,
thus stating that the condition (\ref{cond3}) is indeed
verified and that $(v^x_{3'})'$ is therefore positive. 

\bigskip
\bigskip
\appendix{\large \bf Appendix B: An explicit expression 
for ${\bar V}_s$}
\bigskip

	In this Appendix we provide an explicit
expression of the velocity of the shock wave propagating
towards the right in the limit of $p_3 \rightarrow p_1$
and when the fluid is ideal. This quantity, which is
necessary to calculate the limiting relative velocity
$(\widetilde{v}^x_{12})_{_{2S}}$ in equation
(\ref{capo1}), can be easily computed as
[cf. eq. (\ref{velshock})]

\begin{equation} {\bar V}_s = \frac{ \rho_2^2 W_2^2 v^x_2  +
	|J_{23'}| \sqrt{J^2_{23'} + \rho_2^2 W_2^2 [1 -(v_2^x)^2]}}
         { \rho_2^2 W_2^2 + J^2_{23'} }   \ ,
\end{equation}
where the mass flux $J_{23'}$ is given by [cf. eq. (\ref{flux2})]
\begin{equation}
J^2_{23'}=-\left(\frac{\gamma}{\gamma-1}\right)\frac{p_1 - p_2}
	{h_{3'}(h_{3'}-1)/p_1 - h_2(h_2-1)/p_2 } \ ,
\end{equation}
and where, finally, $h_{3'}$ is the positive root of the
Taub adibat (\ref{Taub2}) in the limit of $p_3
\rightarrow p_1$, i.e.
\begin{equation}
\label{taub_r}
h_{3'} = \frac{(\sqrt{{\cal D}} - 1)(\gamma - 1)(p_1 -
	 p_2)}{2[(\gamma - 1)p_2 + p_1]}
	 \ .
\end{equation}
The quantity ${\cal D}$ in the root (\ref{taub_r}) is
just a shorthand for
\begin{equation}
{\cal D}= 1 - 4 \gamma p_1 \frac{(\gamma-1)p_2 +
	p_1}{(\gamma-1)^2(p_1 - p_2)^2}\left[\frac{h_2(p_2 -
	p_1)}{\rho_2} - h_2^2\right]
 \ .
\end{equation}

\bigskip
\acknowledgements

It is a pleasure to thank J. C. Miller and
J. M$^{\underline{\mbox a}}$ Ib\'a\~nez for useful
discussions and comments. Financial support for this
research has been provided by the Italian MIUR and by the
EU Programme ``Improving the Human Research Potential and
the Socio-Economic Knowledge Base'' (Research Training
Network Contract HPRN-CT-2000-00137).  J.A.P. is
supported by the Marie Curie Fellowship
No. HPMF-CT-2001-01217.

\end{document}